\documentclass[showpacs,showkeys,preprintnumbers]{revtex4}%
\usepackage{amsfonts}
\usepackage{amsmath}
\usepackage{graphicx}
\usepackage{dcolumn}
\usepackage{bm}
\usepackage{textcomp}
\usepackage{amssymb}%
\begin{document}
\title{Phenomenology of charmed mesons in the extended Linear Sigma Model}
\author{Walaa I.\ Eshraim$^{1}$, Francesco Giacosa$^{1,2}$, and Dirk H.\ Rischke$^{1}$}
\affiliation{$^{1}$Institute for Theoretical Physics, Goethe University,
Max-von-Laue-Str.\ 1, D--60438 Frankfurt am Main, Germany }
\affiliation{$^{2}$Institute of Physics, Jan Kochanowski University, 25-406 Kielce, Poland}

\begin{abstract}
We study the so-called extended linear sigma model for the case of four quark
flavors. This model is based on global chiral symmetry and dilatation
invariance and includes, besides scalar and pseudoscalar mesons, vector and
axial-vector mesons. Most of the parameters of the model have been
determined in previous work by fitting properties of mesons with three
quark flavors. Only three new parameters, all related to the current charm
quark mass, appear when introducing charmed mesons. Surprisingly, within the
accuracy expected from our approach, the masses of
open charmed mesons turn out to be in quantitative agreement with
experimental data. On the other hand, with the exception of $J/\psi$,
the masses of charmonia are underpredicted
by about 10\%. It is remarkable
that our approach correctly predicts (within errors)
the mass splitting between spin-0 and spin-1 
negative-parity open charm states. This indicates that, although
the charm quark mass breaks chiral symmetry quite strongly explicitly, 
this symmetry still seems to have some influence on the properties
of charmed mesons.

\end{abstract}

\maketitle

\section{Introduction}

\indent

Quantum chromodynamics (QCD) describes the fundamental interactions of quarks
and gluons. However, due to confinement, in the low-energy domain the physical
degrees of freedom are hadrons, i.e., mesons and baryons. In the last decades,
effective low-energy approaches of the strong interaction have been developed
by imposing chiral symmetry which is one of the basic symmetries of the QCD
Lagrangian in the limit of vanishing quark masses (the so-called chiral limit)
\cite{Gasiorowicz, Meissner}. Chiral symmetry is explicitly broken by the
nonzero current quark masses, but it is also spontaneously broken by a nonzero
quark condensate in the QCD vacuum \cite{Vafa}: as a consequence, pseudoscalar
(quasi-)Goldstone bosons emerge. In a world with only $u$ and $d$ quarks,
i.e., for $N_{f}=2$ quark flavors, these are the pions, while for $N_{f}=3$,
i.e., when also the strange quark $s$ is considered, the Goldstone
modes comprise also kaons and the $\eta$ meson besides the pions. 
[The $\eta^{\prime}$ meson is not a Goldstone
boson because of the chiral anomaly \cite{Pisarski,Hooft}.]

Chiral symmetry can be realized in hadronic approaches in the so-called
nonlinear or in the linear representation. In the nonlinear case, only the
Goldstone bosons are considered \cite{chpt,Schwinger1} [in recent extensions
also vector mesons are added, see e.g.\ Ref.\ \cite{chptvm}]. On the contrary,
in the linear case also the chiral partners of the Goldstone bosons, the
scalar mesons, are retained \cite{Schwinger2, Ko, Urban}. When extending this
approach to the vector sector, both vector and axial-vector mesons are present
\cite{Ko,Urban}. Along this line, recent efforts have led to the construction
of the so-called extended linear sigma model (eLSM), first for $N_{f}=2$
\cite{denis,stani,Gallas} and then for $N_{f}=3$ \cite{dick}. In the eLSM,
besides chiral symmetry, a basic phenomenon of QCD in the chiral limit has
been taken into account: the symmetry under dilatation transformation and its
anomalous breaking (trace anomaly), see e.g.\ Ref.\ \cite{Rosenzweig}. As a
result, the eLSM Lagrangian contains only a \emph{finite} number of terms. In
the case $N_{f}=3$ it was for the first time possible to describe
(pseudo)scalar as well as (axial-)vector meson nonets in a chiral framework
\ \cite{dick}: masses and decay widths turn out to be in very good agreement
with the results listed by the Particle Data Group (PDG) \cite{PDG}.

In this work we investigate the eLSM model in the four-flavor case ($N_{f}%
=4$), i.e., by considering mesons which contain at least one charm quark. This
study is a straightforward extension of Ref.\ \cite{dick}: the Lagrangian has
the same structure as in the $N_{f}=3$ case, except that all (pseudo)scalar
and (axial-)vector meson fields are now parametrized in terms of $4\times4$
(instead of $3\times3$) matrices. These now also include the charmed degrees
of freedom. Since low-energy (i.e., nonstrange and strange) hadron
phenomenology was described very well \cite{dick}, we retain the values for
the parameters that already appear in the three-flavor sector. Then, extending
the model to $N_{f}=4$, three additional parameters enter, all of which are
related to the current charm quark mass (two of them in the (pseudo)scalar
sector and one in the (axial-)vector sector).

Considering that the explicit breaking of chiral and dilatation
symmetries by the current charm quark mass, $m_{c}\simeq1.275$ GeV, is quite
large, one may wonder whether it is at all justified to apply a model based on
chiral symmetry. Rather, one would expect heavy-quark spin symmetry \cite{hqet} to play 
the dominant role. Related to this, the charmed mesons entering our model have a
mass up to about $3.5$ GeV, i.e., they are strictly speaking no longer part of
the low-energy domain of the strong interaction. Naturally, we do not expect
to achieve the same precision as refined potential models \cite{models1,
models2}, lattice-QCD calculations \cite{lattice}, and heavy-quark effective
theories \cite{hqet,bardeen,nowakrho,lutz} [see also the review of
Ref.\ \cite{brambilla} and refs.\ therein]. Nevertheless, it is still
interesting to see how a successful model for low-energy hadron phenomenology
based on chiral symmetry and dilatation invariance fares when extending it to
the high-energy charm sector. Quite surprisingly, a quantitative agreement with
experimental values for the open charmed meson masses is obtained by fitting
just the three additional parameters mentioned above (with deviations of the order of 150 
MeV, i.e., $\sim 5$\%). 
On the other hand, with the exception of $J/\psi$, the charmonium
states turn out to be about 10\% too light
when compared to experimental data. Nevertheless, the main conclusion
is that it is, to first approximation, not unreasonable to delegate the strong breaking
of chiral and dilatation symmetries to three mass terms only and
still have chirally and dilatation invariant interaction terms. Moreover, our
model correctly predicts the mass splitting between spin-0 and spin-1 negative-parity
open charm states, i.e., naturally incorporates the right amount of breaking of 
the heavy-quark spin symmetry.

In addition to meson masses, we study the strong OZI-dominant decays of the
heavy charmed states into light mesons. In this way, our model acts as a
bridge between the high- and low-energy sector of the strong interaction. It
turns out that the OZI-dominant decays are in qualitative agreement with current
experimental results, although the theoretical uncertainties for some of them
are still very large. Nevertheless, since our decay amplitudes depend on the
parameters of the low-energy sector of the theory, there seems to be an
important influence of chiral symmetry in the determination of the decay
widths of charmed states. As a by-product of our analysis, we also obtain the
value of the charm-anticharm condensate and the values of the weak decay
constants of $D$ mesons. Moreover, in the light of our results we shall
discuss the interpretation of the enigmatic scalar strange-charmed meson
$D_{S0}^{\ast}(2317)$ and the axial-vector strange-charmed mesons
$D_{S1}(2460)$ and $D_{S1}(2536)$ \cite{brambilla,lattds,marc,ebert}.

This work is organized as follows: in Sec.\ 2 we present the eLSM for
$N_{f}=4$. In Sec.\ 3 we discuss the results for the masses and decay widths,
and in Sec.\ 4 we give our conclusions. Details of the calculations are
relegated to the Appendices. Our units are $\hbar=c=1$, the metric tensor is
$g^{\mu\nu}=\mathrm{diag}(+,-,-,-)$.


\section{The Lagrangian of the eLSM for $N_{f}=4$}
\label{Lagr}

\indent In this section we extend the eLSM \cite{dick, denisthesis, three
flavor, stani, GiacosaL} to the four-flavor case. To this end, we introduce
$4\times4$ matrices which contain, in addition to the usual nonstrange and
strange mesons, also charmed states. The matrix of pseudoscalar fields $P$
(with quantum numbers $J^{PC}=0^{-+}$) reads
\begin{equation}
P=\frac{1}{\sqrt{2}}\left(
\begin{array}
[c]{cccc}%
\frac{1}{\sqrt{2}}(\eta_{N}+\pi^{0}) & \pi^{+} & K^{+} & D^{0}\\
\pi^{-} & \frac{1}{\sqrt{2}}(\eta_{N}-\pi^{0}) & K^{0} & D^{-}\\
K^{-} & \overline{K}^{0} & \eta_{S} & D_{S}^{-}\\
\overline{D}^{0} & D^{+} & D_{S}^{+} & \eta_{c}%
\end{array}
\right)  \sim\frac{1}{\sqrt{2}}\left(
\begin{array}
[c]{cccc}%
\bar{u}\Gamma u & \bar{d}\Gamma u & \bar{s}\Gamma u & \bar{c}\Gamma u\\
\bar{u}\Gamma d & \bar{d}\Gamma d & \bar{s}\Gamma d & \bar{c}\Gamma d\\
\bar{u}\Gamma s & \bar{d}\Gamma s & \bar{s}\Gamma s & \bar{c}\Gamma s\\
\bar{u}\Gamma c & \bar{d}\Gamma c & \bar{s}\Gamma c & \bar{c}\Gamma c
\end{array}
\right)  \text{ ,} \label{p}%
\end{equation}
where, for sake of clarity, we also show the quark-antiquark content of the
mesons (in the pseudoscalar channel $\Gamma=i\gamma^{5}$). In the
nonstrange-strange sector (the upper left $3\times3$ matrix) the matrix $P$
contains the pion triplet $\vec{\pi},$ the four kaon states $K^{+},$ $K^{-},$
$K^{0},$ $\bar{K}^{0},$ and the isoscalar fields $\eta_{N}=\sqrt{1/2}(\bar
{u}u+\bar{d}d)$ and $\eta_{S}=\bar{s}s.$ The latter two fields mix and
generate the physical fields $\eta$ and $\eta^{\prime}$ [for details, see
Ref.\ \cite{dick}]. In the charm sector (fourth line and fourth column) the
matrix $P$ contains the open charmed states $D^{+},$ $D^{-},$ $D^{0},$
$\bar{D}^{0}$, which correspond to the well-established $D$ resonance, the
open strange-charmed states $D_{S}^{\pm}$, and, finally, the hidden charmed
state $\eta_{c},$ which represents the well-known pseudoscalar ground state
charmonium $\eta_{c}(1S)$.

The matrix of scalar fields $S$ (with quantum numbers $J^{PC}=0^{++}$) reads%
\begin{equation}
S=\frac{1}{\sqrt{2}}\left(
\begin{array}
[c]{cccc}%
\frac{1}{\sqrt{2}}(\sigma_{N}+a_{0}^{0}) & a_{0}^{+} & K_{0}^{\ast+} &
D_{0}^{\ast0}\\
a_{0}^{-} & \frac{1}{\sqrt{2}}(\sigma_{N}-a_{0}^{0}) & K_{0}^{\ast0} &
D_{0}^{\ast-}\\
K_{0}^{\ast-} & \overline{K}_{0}^{\ast0} & \sigma_{S} & D_{S0}^{\ast-}\\
\overline{D}_{0}^{\ast0} & D_{0}^{\ast+} & D_{S0}^{\ast+} & \chi_{c0}%
\end{array}
\right)  \text{ .}%
\end{equation}
The quark-antiquark content is the same as in Eq.\ (\ref{p}), but using
$\Gamma=1_{4}$. A long debate about the correct assignment of light scalar
states has taken place in the last decades. Present results \cite{Amsler,
scalars}, which have been independently confirmed in the framework of the eLSM
\cite{dick}, show that the scalar quarkonia have masses between 1-2 GeV. In
particular, the isotriplet $\vec{a}_{0}$ is assigned to the resonance
$a_{0}(1450)$ (and not to the lighter state $a_{0}(980)$). Similarly, the
kaonic states $K_{0}^{\ast+},$ $K_{0}^{\ast-},$ $K_{0}^{\ast+},$ $\overline
{K}_{0}^{\ast0}$ are assigned to the resonance $K_{0}^{\ast}(1430)$ (and not
to the $K_{0}^{\ast}(800)$ state). The situation in the scalar-isoscalar
sector is more complicated, due to the presence of a scalar glueball state
$G$, see Ref.\ \cite{stani} and below. Then, $\sigma_{N},$ $\sigma_{S}$, $G$
mix and generate the three resonances $f_{0}(1370),$ $f_{0}(1500),$ and
$f_{0}(1710).$ There is evidence \cite{staniproc} that $f_{0}(1370)$ is
predominantly a $\sqrt{1/2}(\bar{u}u+\bar{d}d)$ state, while $f_{0}(1500)$ is
predominantly a $\bar{s}s$ state and $f_{0}(1710)$ predominantly a glueball
state. As a consequence, the light scalar states $f_{0}(500)$ and $f_{0}(980)$
are not quarkonia (but, arguably, tetraquark or molecular states)
\cite{Caprini, Bugg, Black,Giacosatq}. In the open charm sector, we assign the
charmed states $D_{0}^{\ast}$ to the resonances $D_{0}^{\ast}(2400)^{0}$ and
$D_{0}^{\ast}(2400)^{\pm}$ (the latter state has not yet been unambiguously
established). In the strange-charm sector we assign the state $D_{S0}^{\ast
\pm}$ to the only existing candidate $D_{S0}^{\ast}(2317)^{\pm}$; it should,
however, be stressed that the latter state has also been interpreted as a
tetraquark or molecular state because it is too light when compared to
quark-model predictions, see
Refs.\ \cite{models1,brambilla,lattds,marc,ebert,fkguo}. In the next section,
we discuss in more detail the possibility that a heavier, very broad (and
therefore not yet discovered) scalar charmed state exists. In the hidden charm
sector the resonance $\chi_{c0}$ corresponds to the ground-state scalar
charmonium $\chi_{c0}(1P)$.

The matrices $P$ and $S$ are used to construct the matrix $\Phi=S+iP,$ which
transforms under the group $U(4)_{R}\times U(4)_{L}$ as $\Phi\rightarrow
U_{L}\Phi U_{R}^{\dagger}$, where $U_{L}$ and $U_{R}$ are independent
$4\times4$ unitary matrices. Due to this property, the matrix $\Phi$ is used
as a building block for the construction of a Lagrangian which is invariant
under the chiral group $U(4)_{R}\times U(4)_{L}$, see below.

We now turn to the vector sector. The matrix $V^{\mu}$ which includes the
vector degrees of freedom is:%
\begin{equation}
V^{\mu}=\frac{1}{\sqrt{2}}\left(
\begin{array}
[c]{cccc}%
\frac{1}{\sqrt{2}}(\omega_{N}+\rho^{0}) & \rho^{+} & K^{\ast}(892)^{+} &
D^{\ast0}\\
\rho^{-} & \frac{1}{\sqrt{2}}(\omega_{N}-\rho^{0}) & K^{\ast}(892)^{0} &
D^{\ast-}\\
K^{\ast}(892)^{-} & \bar{K}^{\ast}(892)^{0} & \omega_{S} & D_{S}^{\ast-}\\
\overline{D}^{\ast0} & D^{\ast+} & D_{S}^{\ast+} & J/\psi
\end{array}
\right)  ^{\mu}\text{ .}%
\end{equation}
The quark-antiquark content is that shown in Eq.\ (\ref{p}), setting
$\Gamma=\gamma^{\mu}.$ The isotriplet field $\vec{\rho}$ corresponds to the
$\rho$ meson, the four kaonic states correspond to the resonance $K^{\ast
}(892),$ the isoscalar states $\omega_{N}$ and $\omega_{S}$ correspond to the
$\omega$ and $\phi$ mesons, respectively. [No mixing between strange and
nonstrange isoscalars is present in the eLSM; this mixing is small anyway
\cite{klempt}.] In the charm sector, the fields $D^{\ast0},$ $\overline
{D}^{\ast0},D^{\ast+},$ and $D^{\ast-}$ correspond to the vector charmed
resonances $D^{\ast}(2007)^{0}$ and $D^{\ast}(2010)^{\pm}$, respectively,
while the strange-charmed $D_{S}^{\ast\pm}$ corresponds to the resonance
$D_{S}^{\ast\pm}$ (with mass $M_{D_{S}^{\ast\pm}}=(2112.3\pm0.5)$ MeV; note,
however, that the quantum numbers $J^{P}=1^{-}$ are not yet fully
established). Finally, $J/\psi$ is the very well-known lowest vector
charmonium state $J/\psi(1S).$

The matrix $A^{\mu}$ describing the axial-vector degrees of freedom is given
by:%
\begin{equation}
A^{\mu}=\frac{1}{\sqrt{2}}\left(
\begin{array}
[c]{cccc}%
\frac{1}{\sqrt{2}}(f_{1,N}+a_{1}^{0}) & a_{1}^{+} & K_{1}^{+} & D_{1}^{0}\\
a_{1}^{-} & \frac{1}{\sqrt{2}}(f_{1,N}-a_{1}^{0}) & K_{1}^{0} & D_{1}^{-}\\
K_{1}^{-} & \bar{K}_{1}^{0} & f_{1,S} & D_{S1}^{-}\\
\bar{D}_{1}^{0} & D_{1}^{+} & D_{S1}^{+} & \chi_{c1}%
\end{array}
\right)  ^{\mu}\,.
\end{equation}
The quark-antiquark content is that shown in Eq.\ (\ref{p}), setting
$\Gamma=\gamma^{\mu}\gamma^{5}.$ The isotriplet field $\vec{a}_{1}$
corresponds to the field $a_{1}(1260),$ the four kaonic states $K_{1}$
correspond (predominantly) to the resonance $K_{1}(1200)$ [but also to
$K_{1}(1400),$ because of mixing between axial-vector and pseudovector states,
see Ref.\ \cite{florianlisa} and refs.\ therein]. The isoscalar fields
$f_{1,N}$ and $f_{1,S}$ correspond to $f_{1}(1285)$ and $f_{1}(1420),$
respectively. In the charm sector, the $D_{1}$ field is chosen to correspond
to the resonances $D_{1}(2420)^{0}$ and $D_{1}(2420)^{\pm}$. (Another
possibility would be the not yet very well established resonance
$D_{1}(2430)^{0}$, or, due to mixing between axial- and pseudovector states,
to a mixture of $D_{1}(2420)$ and $D_{1}(2430)$. Irrespective of this
uncertainty, the small mass difference between these states would leave our
results virtually unchanged.) The assignment of the strange-charmed doublet
$D_{S1}^{\pm}$ is not yet settled, the two possibilities listed by the PDG are
the resonances $D_{S1}(2460)^{\pm}$ and $D_{S1}(2536)^{\pm}$ \cite{PDG}.
According to various studies, the latter option is favored, while the former
can be interpreted as a molecular or a tetraquark state \cite{fkguo,
Ds1molecule, Ds vector}. Thus, we assign our quark-antiquark $D_{1}$ state to
the resonance $D_{S1}(2536)^{\pm}.$ Finally, the charm-anticharm state
$\chi_{c1}$ can be unambiguously assigned to the charm-anticharm resonance
$\chi_{c1}(1P)$.

From the matrices $V^{\mu}$ and $A^{\mu}$ we construct the left-handed and
right-handed vector fields $L^{\mu}=V^{\mu}+A^{\mu}$ and $R^{\mu}=V^{\mu
}-A^{\mu}$. Under chiral transformations they transform as $L^{\mu}\rightarrow
U_{L}L^{\mu}U_{L}^{\dag}$ and $R^{\mu}\rightarrow U_{R}L^{\mu}U_{R}^{\dag}.$
Due to this transformation property they are used as building blocks of the
chirally invariant Lagrangian, see below.

The last field entering the model is the scalar glueball $G$, described by the
dilaton Lagrangian
\begin{equation}
\mathcal{L}_{dil}=\frac{1}{2}(\partial_{\mu}G)^{2}-\frac{1}{4}\frac{m_{G}^{2}%
}{\Lambda^{2}}\left(  G^{4}\,\log\frac{G}{\Lambda}-\frac{G^{4}}{4}\right)
\text{ ,} \label{dil}%
\end{equation}
which mimics the trace anomaly of QCD \cite{Rosenzweig, dick, nf3Giacosa}. The
dimensionful parameter $\Lambda$ sets the energy scale of low-energy QCD; in
the chiral limit it is the only dimensionful parameter besides the coefficient
of the term representing the axial anomaly. All other interaction terms of the
Lagrangian are described by dimensionless coupling constants. The parameter
$m_{G}$ is the glueball mass in the quenched approximation (no quarks), which
is about 1.5-1.7 GeV \cite{Morningstar}. As mentioned above, the
identification of $G$ is still uncertain, the two most likely candidates are
$f_{0}(1500)$ and $f_{0}(1710)$ and/or admixtures of them. Note that, while we
include the scalar glueball because it is conceptually important to guarantee
dilatation invariance of the model (thus constraining the number of possible
terms that it can have), just as in Ref.\ \cite{dick} we do not make an
assignment for the glueball, since it does not affect the results of the present study. [The
pseudoscalar glueball $\tilde{G}$ has been also added to the eLSM
\cite{Eshraim}, but it is also not relevant here and therefore omitted.]

The full Lagrangian is constructed by requiring chiral symmetry, dilatation
invariance (besides terms related to the current quark masses, the $U(1)_{A}$
anomaly, and the dilaton potential, which break this symmetry explicitly), and
invariance under the discrete symmetries $C$ and $P$:
\begin{align}
\mathcal{L}  &  =\mathcal{L}_{dil}+\mathrm{Tr}[(D^{\mu}\Phi)^{\dagger}(D^{\mu
}\Phi)]-m_{0}^{2}\left(  \frac{G}{G_{0}}\right)  ^{2}\mathrm{Tr}(\Phi
^{\dagger}\Phi)-\lambda_{1}[\mathrm{Tr}(\Phi^{\dagger}\Phi)]^{2} -\lambda
_{2}\mathrm{Tr}(\Phi^{\dagger}\Phi)^{2}+\mathrm{Tr}[H(\Phi+\Phi^{\dagger
})]\nonumber\\
&  +\mathrm{Tr}\left\{  \left[  \left(  \frac{G}{G_{0}}\right)  ^{2}%
\frac{m_{1}^{2}}{2}+\Delta\right]  \left[  (L^{\mu})^{2}+(R^{\mu})^{2}\right]
\right\}  -\frac{1}{4}\mathrm{Tr}[(L^{\mu\nu})^{2}+(R^{\mu\nu})^{2}%
]-2\,\mathrm{Tr}[\varepsilon\Phi^{\dagger}\Phi] +c(\mathrm{det}\Phi
-\mathrm{det}\Phi^{\dagger})^{2}\nonumber\\
&  +i\frac{g_{2}}{2}\{\mathrm{Tr}(L_{\mu\nu}[L^{\mu},L^{\nu}])+\mathrm{Tr}%
(R_{\mu\nu}[R^{\mu},R^{\nu}])\}+\frac{h_{1}}{2}\mathrm{Tr}(\Phi^{\dagger}%
\Phi)\mathrm{Tr}[(L^{\mu})^{2}+(R^{\mu})^{2}]+h_{2}\mathrm{Tr}[|\Phi R^{\mu
}|^{2}+|L^{\mu}\Phi|^{2}]\nonumber\\
&  +2h_{3}\mathrm{Tr}(\Phi R_{\mu}\Phi^{\dagger}L^{\mu})+ \ldots\,,\,
\label{lag}%
\end{align}
where $\mathcal{L}_{dil}$ is the dilaton Lagrangian of Eq.\ (\ref{dil}),
$D^{\mu}\Phi\equiv\partial^{\mu}\Phi-ig_{1}(L^{\mu}\Phi-\Phi R^{\mu})$,
$L^{\mu\nu}\equiv\partial^{\mu}L^{\nu}-\partial^{\nu}L^{\mu}$, and $R^{\mu\nu
}\equiv\partial^{\mu}R^{\nu}-\partial^{\nu}R^{\mu}$. The dots refer to further
chirally invariant terms listed in Ref.\ \cite{dick}; these terms do not
affect the masses and decay widths studied in this work and are therefore omitted.

The terms involving the matrices $H,\varepsilon,$ and $\Delta$ describe the
breaking of dilatation and chiral symmetry due to nonzero current quark
masses. They are of particular importance when the charmed mesons are
considered, because the charm quark mass is large. We describe them separately:

(i) The term $\mathrm{Tr}[H(\Phi+\Phi^{\dagger})]$ with%

\begin{equation}
H=\frac{1}{2}\left(
\begin{array}
[c]{cccc}%
h_{U} & 0 & 0 & 0\\
0 & h_{D} & 0 & 0\\
0 & 0 & \sqrt{2}h_{S} & 0\\
0 & 0 & 0 & \sqrt{2}h_{C}%
\end{array}
\right)  \, , \label{h}%
\end{equation}
describes the usual explicit symmetry breaking (tilting of the Mexican-hat
potential). The constants $h_{i}$ are proportional to the current quark
masses, $h_{i}\propto m_{i}.$ Here we work in the isospin-symmetric limit,
$h_{U}=h_{D}=h_{N}.$ The pion mass, for instance, turns out to be $m_{\pi}%
^{2}\propto m_{u},$ in agreement with the Gell-Mann--Oakes--Renner (GOR)
relation \cite{gor}. The parameter $h_{C}$ is one of the three new parameters
entering the $N_{f}=4$ version of the model when compared to the $N_{f}=3$
case of Ref.\ \cite{dick}.

(ii) The term $-2\,\mathrm{Tr}[\varepsilon\Phi^{\dagger}\Phi]$ with
\begin{equation}
\varepsilon=\left(
\begin{array}
[c]{cccc}%
\varepsilon_{U} & 0 & 0 & 0\\
0 & \varepsilon_{D} & 0 & 0\\
0 & 0 & \varepsilon_{S} & 0\\
0 & 0 & 0 & \varepsilon_{C}%
\end{array}
\right)  \text{ ,} \label{8}%
\end{equation}
where $\varepsilon_{i}\propto m_{i}^{2}$ is the direct
contribution of the bare quark masses to the (pseudo)scalar meson masses. 
In the isospin-symmetric limit
$\varepsilon_{U}=\varepsilon_{D}=\varepsilon_{N}$ one can subtract from
$\varepsilon$ a matrix proportional to the identity in such a way that the
parameter $\varepsilon_{N}$ can be absorbed in the parameter $m_{0}^{2}.$
Thus, without loss of generality we can set $\varepsilon_{N}=0.$ Following
Ref.\ \cite{dick}, for the sake of simplicity we shall here also set
$\varepsilon_{S}=0,$ while we keep $\varepsilon_{C}$ nonzero. This is the
second additional parameter with respect to Ref.\ \cite{dick}.

(iii) The term $\mathrm{Tr}\left[  \Delta(L^{\mu}{}^{2}+R^{\mu}{}^{2})\right]
$ with
\begin{equation}
\Delta=\left(
\begin{array}
[c]{cccc}%
\delta_{U} & 0 & 0 & 0\\
0 & \delta_{D} & 0 & 0\\
0 & 0 & \delta_{S} & 0\\
0 & 0 & 0 & \delta_{C}%
\end{array}
\right)  \,, \label{7}%
\end{equation}
where $\delta_{i}\sim m_{i}^{2}$ describes the current quark-mass
contribution to the masses of the (axial-)vector mesons. Also in this case, in
the isospin-symmetric limit it is possible to set $\delta_{U}=\delta
_{D}=\delta_{N}=0$ because an identity matrix can be absorbed in the term
proportional to $m_{1}^{2}.$ The parameter $\delta_{S}$ is taken from
Ref.\ \cite{dick}. The third new parameter with respect to Ref.\ \cite{dick}
is $\delta_{C}.$

Note that in the present effective model the mass parameters
$\delta_{C}$ and $\varepsilon_{C}$ are not to be regarded as the
second-order contribution in an expansion in powers of $m_{c}$. They simply represent
the direct, and in this case dominant, contribution $\sim m_c$ of a charm quark 
to the masses of charmed (pseudo)scalar and (axial-)vector mesons.

Another important term in the Lagrangian (\ref{lag}) is $c(\mathrm{det}%
\Phi-\mathrm{det}\Phi^{\dagger})^{2},$ which describes the axial anomaly and
is responsible for the large $\eta^{\prime}$ mass. Care is needed, because the
determinant changes when the number of flavors changes. The relation between
$c$ and its counterpart in the three-flavor case $c_{N_{f}=3}$ of
Ref.\ \cite{dick} is given by:
\begin{equation}
c=\frac{2\,c_{N_{f}=3}}{\phi_{C}^{2}}\text{ .} \label{cc1}%
\end{equation}
Thus, the parameter $c$ can be determined once the condensate $\phi_{C}$ is
obtained, see next section.

The Lagrangian (\ref{lag}) induces spontaneous symmetry breaking if $m_{0}%
^{2}<0$: as a consequence, the scalar-isoscalar fields $G,\sigma_{N}%
,\,\sigma_{S},$ and $\chi_{c0}$ develop nonzero vacuum expectation values. One
has to perform the shifts $G\rightarrow G+G_{0}$, $\sigma_{N}\rightarrow
\sigma_{N}+\phi_{N}\,\,,$ $\sigma_{S}\rightarrow\sigma_{S}+\phi_{S},$ and
$\chi_{c0}\rightarrow\chi_{c0}+\phi_{C}.$ The quantity $G_{0}$ is proportional
to the gluon condensate \cite{stani}, while $\phi_{N},$ $\phi_{S},$ and
$\phi_{C}$ correspond to the nonstrange, strange, and charm quark-antiquark condensates.

Once the shifts are performed, one also has to get rid of mixing between
(axial-)vector and (pseudo)scalar states by a proper shift of axial-vector
fields and renormalization of the (pseudo)scalar ones. Then, the physical
masses and interaction terms can be calculated. We present the detailed
expressions for the masses in App.\ A.

We conclude this section with a few remarks on how to extend
the Lagrangian (\ref{lag}) in order to improve the description of hadron vacuum
properties. First of all, note that the requirement of dilatation invariance restricts 
the interaction terms in the Lagrangian to have naive scaling dimension four: 
higher-order dilatation-invariant terms would have to contain inverse powers of $G$, 
and thus would be non-analytic in this field. In this sense, our Lagrangian is
complete and cannot be systematically improved by the inclusion
of higher-order interaction terms, such as in theories with nonlinearly realized
chiral symmetry. However, one may add further terms that violate dilatation
invariance (which is already broken by the mass terms $\sim H, \varepsilon, \Delta$,
and the $U(1)_A$--violating term $\sim c$) to improve the model. 

Another possibility is to sacrifice chiral symmetry. For instance, since the
explicit breaking of chiral symmetry by the charm quark mass is large (which
is accounted for by the terms $\sim h_C, \varepsilon_C, \delta_C$), one could
also consider chiral-symmetry violating interaction terms, e.g.\ replace
\begin{equation} \label{U4symbreak}
\lambda_{2}\mathrm{Tr}(\Phi^{\dagger}  \Phi)^{2}
\longrightarrow \lambda_{2}\mathrm{Tr}(\Phi^{\dagger}  \Phi)^{2}
+ \delta \lambda_2 \mathrm{Tr}( \mathbb{P}_C \Phi^{\dagger} \Phi)^{2}
\end{equation}
where $\mathbb{P}_C = {\rm diag}\{0,0,0,1\}$ is a projection operator onto the
charmed states. A value $\delta\lambda_{2}\neq 0$ 
explicitly breaks the symmetry of this interaction term from 
$U_{R}(4)\times U_{L}(4)$ to $U_{R}(3)\times U_{L}(3)$. 
One could modify the interaction terms proportional to $\lambda_1$,
$c$, $g_{1}$, $g_{2},$ $h_{1},$ $h_{2},$
and $h_{3}$ in Eq.\ (\ref{lag}) in a similar manner.

\section{Results}

\subsection{Masses}

The Lagrangian (\ref{lag}) contains the following 15 free parameters:
$m_{0}^{2},\,\lambda_{1},\,\lambda_{2},\,m_{1},\,g_{1},\,c_{1},\,h_{1}%
,\,h_{2},\,h_{3},\,\delta_{S},\,\delta_{C},\,\varepsilon_{C},$ $h_{N},$
$h_{S},$ and $h_{C}$. For technical reasons, instead of the parameters
$h_{N},$ $h_{S},$ and $h_{C}$ entering Eq.\ (\ref{h}), it is easier to use the
condensates $\phi_{N},$ $\phi_{S},$ $\phi_{C}$. This is obviously equivalent,
because $\phi_{N},$ $\phi_{S},$ and $\phi_{C}$ form linearly independent
combinations of the parameters.

In the large-$N_{c}$ limit, one sets $h_{1}=\lambda_{1}=0.$ Then, as shown in
Ref.\ \cite{dick} for the case $N_{f}=3$, ten parameters can be determined by
a fit to masses and decay widths of mesons below 1.5 GeV as shown in Tab.\ 1.
In the following we use these values for our numerical calculations. As a
consequence, the masses and the decay widths of the nonstrange-strange mesons
are -- by construction -- identical to the results of Ref.\ \cite{dick} (see
Tab.\ 2 and Fig.\ 1 in that reference). Note also that, in virtue of
Eq.\ (\ref{cc1}), the parameter combination $\phi_{C}^{2}c/2$ is determined by
the fit of Ref.\ \cite{dick}.

\begin{center}
\textbf{Table\thinspace1}\thinspace:\thinspace\ Values of the parameters [from
Ref.\ \cite{dick}].%

\begin{tabular}
[c]{|c|c|c|c|}\hline
Parameter & Value & Parameter & Value\\\hline
$m_{1}^{2}$ & $0.413\times10^{6}$ MeV$^{2}$ & $m_{0}^{2}$ & $-0.918\times
10^{6}$ MeV$^{2}$\\\hline
$\phi_{C}^{2}c/2$ & $450\times10^{-6}$ MeV$^{-2}$ & $\delta_{S}$ &
$0.151\times10^{6}$MeV$^{2}$\\\hline
$g_{1}$ & $5.84$ & $h_{1}$ & $0$\\\hline
$h_{2}$ & $9.88$ & $h_{3}$ & $3.87$\\\hline
$\phi_{N}$ & $164.6$ MeV & $\phi_{S}$ & $126.2$ MeV\\\hline
$\lambda_{1}$ & $0$ & $\lambda_{2}$ & $68.3$\\\hline
\end{tabular}

\end{center}

For the purposes of the present work, we are left with three unknown
parameters: $\phi_{C},\varepsilon_{C},$ $\delta_{C}.$ We determine them by
performing a fit to twelve experimental (hidden and open) charmed meson masses
listed by the PDG \cite{PDG}
\begin{equation}
\chi^{2}\equiv\sum_{i}^{12}\bigg(\frac{M_{i}^{th}-M_{i}^{exp}}{\xi M_{i}%
^{exp}}\bigg)^{2}\text{ ,} \label{chi2}%
\end{equation}
where $\xi$ is a constant. We do not use the experimental errors for the
masses, because we do not expect to reach the same precision with our
effective model which, besides other effects, already neglects isospin
breaking. In Ref.\ \cite{dick}, we required a minimum error of 5\% for
experimental quantities entering our fit, and obtained a reduced $\chi^{2}$ of
about 1.23. Here, we slightly change our fit strategy: we choose the parameter
$\xi$ such that the reduced $\chi^{2}$ takes the value $\chi^{2}/(12-3)=1,$
which yields $\xi=0.07.$ This implies that we enlarge the experimental errors
to $7\%$ of the respective masses.

The parameters (together with their theoretical errors) are:
\begin{equation}
\phi_{C}=(176\pm28)\text{ MeV},\,\,\,\delta_{C}=(3.91\pm0.36)\times
10^{6}\text{ MeV}^{2},\text{ }\varepsilon_{C}=(2.23\pm0.71)\times10^{6}\text{
MeV}^{2}\text{ .} \label{fit}%
\end{equation}

In Tab.\ 2 we present the results of our fit by comparing the theoretically
computed with the experimentally measured masses [see also
Ref.\ \cite{EshraimProc1} for preliminary results]. For the nonstrange-charmed
states we use the masses of the neutral members of the multiplet in the fit,
because the corresponding resonances have been clearly identified and the
masses have been well determined for all quantum numbers. In view of the fact
that the employed model is built as a low-energy chiral model and that only
three parameters enter the fit, the masses are quite well described. The
mismatch grows for increasing masses because Eq.\ (\ref{chi2}) imposes, by
construction, a better precision for low masses. For comparison, in the right
column of Tab.\ 2 we also show the value $0.07M_{i}^{exp}$ which represents
the `artificial experimental error' that we have used in our fit.

\begin{center}
\textbf{Table\thinspace2}\thinspace:\thinspace\ Masses of charmed meson used
in the fit.%

\begin{tabular}
[c]{|c|c|c|c|c|c|}\hline
Resonance & Quark content & $J^{P}$ & Our Value [MeV] & Experimental Value
[MeV] & 7\% of the exp. value [MeV]\\\hline
$D^{0}$ & $u\bar{c},\bar{u}c$ & $0^{-}$ & $1981\pm73$ & $1864.86\pm0.13$ &
$130$\\\hline
$D_{S}^{\pm}$ & $s\bar{c},\bar{s}c$ & $0^{-}$ & $2004\pm74$ & $1968.50\pm0.32$
& $138$\\\hline
$\eta_{c}(1S)$ & $c\bar{c}$ & $0^{-}$ & $2673\pm118$ & $2983.7\pm0.7$ &
$209$\\\hline
$D_{0}^{\ast}(2400)^{0}$ & $u\bar{c},\bar{u}c$ & $0^{+}$ & $2414\pm77$ &
$2318\pm29$ & $162$\\\hline
$D_{S0}^{\ast}(2317)^{\pm}$ & $s\bar{c},\bar{s}c$ & $0^{+}$ & $2467\pm76$ &
$2317.8\pm0.6$ & $162$\\\hline
$\chi_{c0}(1P)$ & $c\bar{c}$ & $0^{+}$ & $3144\pm128$ & $3414.75\pm0.31$ &
$239$\\\hline
$D^{\ast}(2007)^{0}$ & $u\bar{c},\bar{u}c$ & $1^{-}$ & $2168\pm70$ &
$2006.99\pm0.15$ & $140$\\\hline
$D_{S}^{\ast}$ & $s\bar{c},\bar{s}c$ & $1^{-}$ & $2203\pm69$ & $2112.3\pm0.5$
& $148$\\\hline
$J/\psi(1S)$ & $c\bar{c}$ & $1^{-}$ & $2947\pm109$ & $3096.916\pm0.011$ &
$217$\\\hline
$D_{1}(2420)^{0}$ & $u\bar{c},\bar{u}c$ & $1^{+}$ & $2429\pm63$ &
$2421.4\pm0.6$ & $169$\\\hline
$D_{S1}(2536)^{\pm}$ & $s\bar{c},\bar{s}c$ & $1^{+}$ & $2480\pm63$ &
$2535.12\pm0.13$ & $177$\\\hline
$\chi_{c1}(1P)$ & $c\bar{c}$ & $1^{+}$ & $3239\pm101$ & $3510.66\pm0.07$ &
$246$\\\hline
\end{tabular}

\end{center}

The following remarks about our results are in order.

(i) Remembering that our model is a low-energy effective approach
to the strong interaction, it is quite surprising that the masses of the open charmed
states are in good quantitative agreement (within the theoretical error) with 
experimental data. In
particular, when taking into account the 7\% range (right column of Tab.\ 2),
almost all the results are within 1$\sigma$ or only slightly above it.
Clearly, our results cannot compete with the precision of other approaches,
but show that a connection to low-energy models is possible.

(ii) With the exception of $J/\psi$, the masses of the charmonia states 
are underestimated by about 10\% as is
particularly visible for the resonance $\eta_{c}(1S).$ On the one hand, this
is due to the way the fit has been performed; on the other hand, it points to
the fact that unique values of the parameters $h_{C},$ $\delta_{C}$, and
$\varepsilon_{C}$ are not sufficient for a precise description of both open
and hidden charmed states over the whole energy range. One way to
improve the fit of the charmonium masses
would be to include non-zero values for $\lambda_1, h_1$.
Another way is explicitly breaking the chiral symmetry as
discussed at the end of Sec.\ \ref{Lagr}. However, since the description
of open charm states is already reasonable, we only need to consider
the charmonium states. For the (pseudo)scalar charmonia, we would 
thus modify the
second term in Eq.\ (\ref{U4symbreak}) by introducing another projection
operator under the trace, ${\rm Tr} (\mathbb{P}_C \Phi^\dagger \Phi)^2
\rightarrow {\rm Tr} (\mathbb{P}_C \Phi^\dagger \mathbb{P}_C \Phi)^2$.
A similar consideration could be done for the (axial-)vector charmonia.

(iii) The experimental value for the mass of the charged scalar state
$D_{0}^{\ast}(2400)^{\pm}$, which is $(2403\pm14\pm35)$ MeV, is in fair
agreement with our theoretical result, although the existence of this
resonance has not yet been unambiguously established.

(iv) The theoretically computed mass of the strange-charmed scalar state
$D_{S0}^{\ast\pm}$ turns out to be larger than that of the charmed state
$D_{0}^{\ast}(2400)^{0}.$ This is a natural consequence of the strange quark
being heavier than up and down quark. 
Note that, however, the experimentally measured masses of 
$D_{S0}^{\ast}(2317)^{\pm}$ and $D_{0}^{\ast}(2400)^{0}$ are
virtually identical, which is somewhat surprising.

(v) The theoretical mass of the axial-vector strange-charmed state $D_{S1}$
reads $2480$ MeV, which lies in between the two physical states $D_{S1}%
(2460)^{\pm}$ and $D_{S1}(2536)^{\pm}.$ We shall re-analyze the scalar and the
axial-vector strange-charmed states in light of the results for the decay
widths, see next subsection.

As a next step, we turn our attention to some mass differences and to the role
of the charm condensate $\phi_{C}.$\textbf{ }Namely, in the following mass
differences the parameters $\varepsilon_{C}$ and $\delta_{C}$ cancels:
\begin{equation}
\,\,\,\,m_{D_{1}}^{2}-m_{D^{\ast}}^{2}=\sqrt{2}\,(g_{1}^{2}-h_{3})\phi
_{N}\,\phi_{C},\,\,\,\,\,\,m_{\chi_{c1}}^{2}-m_{J/\psi}^{2}=2\,(g_{1}%
^{2}-h_{3})\phi_{C}^{2},\text{ }m_{D_{S1}}^{2}-m_{D_{S}^{\ast}}^{2}%
=2\,(g_{1}^{2}-h_{3})\phi_{S}\,\phi_{C}\,.\label{massdiff}%
\end{equation}
Using Eq.\ (\ref{fit}), the theoretical results are
\[
m_{D_{1}}^{2}-m_{D^{\ast}}^{2}=(1.2\pm0.6)\times10^{6}\text{ MeV}^{2}\text{ ,
}\,\,\,m_{\chi_{c1}}^{2}-m_{J/\psi}^{2}=(1.8\pm1.3)\times10^{6}\text{ MeV}%
^{2}\text{, }\,\,\,m_{D_{S1}}^{2}-m_{D_{S}^{\ast}}^{2}=(1.2\pm0.6)\times
10^{6}\text{ MeV}^{2}\,,
\]
while the experimental values are (the experimental errors are omitted
because, being of the order of $10^{3}$ MeV$^{2},$ they are very small when
compared to the theoretical ones):
\[
m_{D_{1}}^{2}-m_{D^{\ast}}^{2}=1.82\times10^{6}\text{ MeV}^{2}\text{ ,
}\,\,\,\,m_{\chi_{c1}}^{2}-m_{J/\psi}^{2}=2.73\times10^{6}\text{ MeV}%
^{2}\text{ , }\,\,\,\,m_{D_{S1}}^{2}-m_{D_{S}^{\ast}}^{2}=1.97\times
10^{6}\text{ MeV}^{2}\,.
\]
The agreement is fairly good, which shows that our determination of the charm
condensate $\phi_{C}$ is compatible with the experiment, although it still has
a large uncertainty. Note that a similar determination of $\phi_{C}$ 
via the weak decay constants of
charmed mesons determined via the PCAC relations has been presented in Ref.\
\cite{mishra}. Their result is $\phi_{C}/\phi_{N}\simeq1.35$, which is compatible to our ratio of
about $1.07\pm0.20$. Previously, Ref.\ \cite{gottfriedklevansky}
determined $\phi_{C}/\phi_{N}\simeq1.08$ in the framework of the NJL model, which is
in perfect agreement with our result.

From a theoretical point of view, it is instructive to study the behavior of
the condensate $\phi_{C}$ as function of the heavy quark mass $m_{c}$. To this
end, we recall that the equation determining $\phi_{C}$ is of the third-order
type and reads (for $\lambda_{1}=0$)%
\begin{equation}
h_{0,C}=(m_{0}^{2}+2\varepsilon_{C})\phi_{C}+\lambda_{2}\phi_{C}^{3}\text{ .}%
\end{equation}
By imposing the scaling behaviors $h_{0,C}=m_{c}\tilde{h}_{0,C}$ and
$\varepsilon_{C}=m_{c}^{2}\tilde{\varepsilon}_{C}$, the solution for large
values of $m_{c}$ reads $\phi_{C}\simeq h_{0,C}/2\varepsilon_{C}\propto
1/m_{c}$, which shows that the mass differences of Eq.\ (\ref{massdiff})
vanish in the heavy-quark limit. However, the fact that the value of the charm
condensate turns out to be quite large implies that the charm quark is not yet
that close to the heavy-quark limit. To show this aspect, we plot in Fig.\ 1
the condensate $\phi_{C}$ as function of $m_{c}$ by keeping all other
parameters fixed. Obviously, this is a simplifying assumption, but the main
point of this study is a qualitative demonstration that the chiral condensate
of charm-anticharm quarks is non-negligible.

Note that, in contrast to the demands of heavy-quark spin symmetry \cite{hqet},
our chiral approach does not necessarily imply an equal mass of the
vector and pseudoscalar states and of scalar and axial-vector states. Namely,
for a very large heavy quark mass one has $m_{D}^{2}\simeq\varepsilon_{C}$ and
$m_{D^{\ast}}^{2}\simeq\delta_{C}.$ Thus, in order to obtain a degeneracy of
these mesons, as predicted by heavy-quark symmetry, we should
additionally impose that $\varepsilon_{C}=\delta_{C}.$ Numerically, the values
have indeed the same order of magnitude, but differ by a factor of two, see
Eq.\ (\ref{fit}). Nevertheless, the difference of masses $m_{D^{\ast}}%
-m_{D}\simeq 180$ MeV of the spin-0 and spin-1 negative-parity open charm
states turns out to
be (at least within theoretical errors) in quantitative agreement with the experiment. 
Thus, at least for negative-parity mesons, our chiral approach seems 
to correctly predict the amount of breaking of the 
heavy-quark spin symmetry. For the mass difference $m_{D_1}-m_{D_0^*}$
of spin-0 and spin-1 mesons with positive parity, our model underpredicts
the experimental values by an order of magnitude, i.e., our approach based
on chiral symmetry follows the predictions of heavy-quark symmetry even more
closely than nature! 

We conclude our discussion of the charmed meson masses by remarking
that chiral symmetry (and its breaking) may still have a sizable influence
in the charm sector. In this context it is interesting to note that in the 
works \cite{nowakrho,bardeen} the degeneracy of vector and axial-vector
charmed states in the heavy-quark limit [see Eq.\ (\ref{massdiff})] as well as
of scalar and pseudoscalar charmed states was obtained by combining the
heavy-quark spin and the (linear) chiral symmetries.

\begin{figure}[Hb]
\begin{center}
\includegraphics[
height=2.9637in,
width=4.9191in]%
{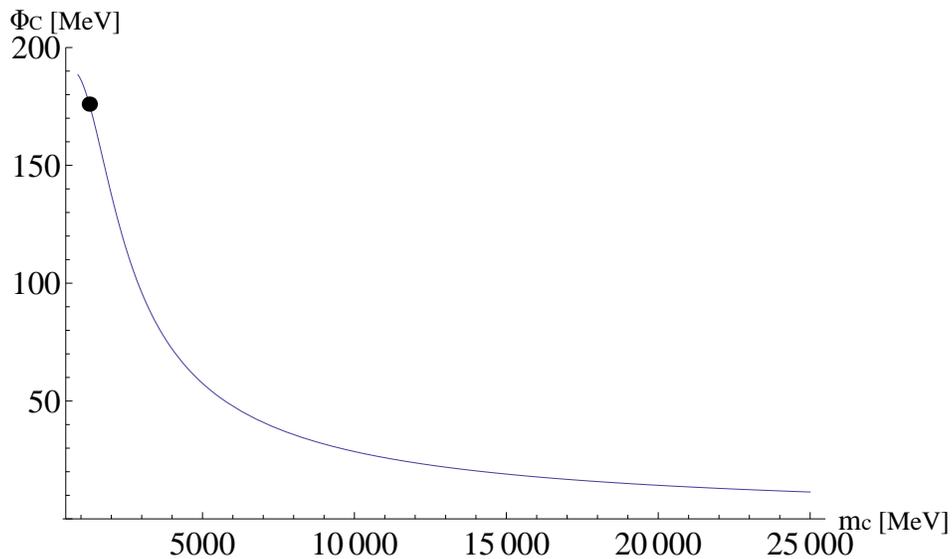}%
\caption{Condensate $\phi_{C}$ as function of the quark mass $m_{c}$. The dot
corresponds to the physical value $m_{c}=1.275$ GeV \cite{PDG}.}%
\end{center}
\end{figure}
EndExpansion

\subsection{Decays}

In this subsection we study the decays of charmed mesons. As a first step we
evaluate the weak-decay constants of the pseudoscalar mesons $D$ and $D_{S}.$
Their analytic expressions read [see App.\ A and also
Ref.\ \cite{EshraimProc1}]%
\begin{equation}
f_{D}=\frac{\phi_{N}+\sqrt{2}\phi_{C}}{\sqrt{2}Z_{D}}\text{ },\text{ }%
f_{D_{S}}=\frac{\phi_{S}+\phi_{C}}{Z_{D_{S}}}\text{ },\text{ }f_{\eta_{c}%
}=\frac{2\phi_{C}}{Z_{\eta_{C}}}\text{ }.
\end{equation}
Using the parameters of the fit we obtain
\begin{equation}
f_{D}=(254\pm17)\text{ MeV },\text{ }f_{D_{S}}=(261\pm17)\text{ MeV },\text{
}f_{\eta_{C}}=(314\pm39)\text{ MeV.}%
\end{equation}
The experimental values $f_{D}=(206.7\pm8.9)$ MeV and $f_{D_{S}}%
=(260.5\pm5.4)$ MeV \cite{PDG} show a good agreement for $f_{D_{S}}$ and a
slightly too large theoretical result for $f_{D}$. The quantity $f_{\eta_{C}}$
is in fair agreement with the experimental value $f_{\eta_{C}}=(335\pm75)$ 
MeV \cite{etacexp} as well as with the theoretical
result $f_{\eta_{C}}=(300\pm50)$ MeV obtained in Ref.\ \cite{etacth2}. 
These results show that our determination of the condensate $\phi_{C}$ is
reliable (even if the theoretical uncertainty is still large).

We now turn to the (OZI-dominant) strong decay widths of the resonances
$D_{0},$ $D^{\ast},$ $D_{1}$, and $D_{S1}$. The results are summarized in
Tab.\ 3. For the calculation of the decay widths we have used the physical
masses listed by the PDG. This is necessary in order to have the correct phase
space. Although the theoretical uncertainties are large and some experimental
results are not yet well known, the qualitative agreement is 
remarkable if one considers that the decay amplitudes depend on the
parameters of the three-flavor version of the model determined in
Ref.\ \cite{dick}. Note that the theoretical errors have been calculated by
taking into account the uncertainty in the charm condensate $\phi_{C}%
=(176\pm28)$ MeV. The lower theoretical value corresponds to $\phi_{C}=\left(
176-28\right)  $ MeV, while the upper one to $\phi_{C}=\left(  176+28\right)
$ MeV. The explicit expressions for the decay widths are reported in App.\ B.

Here we do not study the decay of other (hidden and open) charmed states
because we restrict ourselves to OZI-dominant processes. The study of
OZI-suppressed decays which involve the large-$N_{c}$ suppressed parameters
$\lambda_{1}$ and $h_{1}$ is left for future work. Then, also the decays of
the well-known charmonium states (such as $\chi_{c0}$ and $\eta_{c}$) will be investigated.

\begin{center}
\textbf{Table\thinspace3:}\thinspace\thinspace\ Decay widths of charmed
mesons
%

\begin{tabular}
[c]{|c|c|c|}\hline
Decay Channel & Theoretical result [MeV] & Experimental result [MeV]\\\hline
$D_{0}^{\ast}(2400)^{0}\rightarrow D\pi=D^{+}\pi^{-}+D^{0}\pi^{0}$ &
$139_{-114}^{+243}$ & $D^{+}\pi^{-}$ seen; full width $\Gamma=267\pm
40$\\\hline
$D_{0}^{\ast}(2400)^{+}\rightarrow D\pi=D^{+}\pi^{0}+D^{0}\pi^{+}$ &
$51_{-51}^{+182}$ & $D^{+}\pi^{0}$ seen; full width: $\Gamma=283\pm24\pm
34$\\\hline
$D^{\ast}(2007)^{0}\rightarrow D^{0}\pi^{0}$ & $0.025\pm0.003$ & seen;
$<1.3$\\\hline
$D^{\ast}(2007)^{0}\rightarrow D^{+}\pi^{-}$ & $0$ & not seen\\\hline
$D^{\ast}(2010)^{+}\rightarrow D^{+}\pi^{0}$ & $0.018_{-0.003}^{+0.002}$ &
$0.029\pm0.008$\\\hline
$D^{\ast}(2010)^{+}\rightarrow D^{0}\pi^{+}$ & $0.038_{-0.004}^{+0.005}$ &
$0.065\pm0.017$\\\hline
$D_{1}(2420)^{0}\rightarrow D^{\ast}\pi=D^{\ast+}\pi^{-}+D^{\ast0}\pi^{0}$ &
$65_{-37}^{+51}$ & $D^{\ast+}\pi^{-}$ seen; full width: $\Gamma=27.4\pm
2.5$\\\hline
$D_{1}(2420)^{0}\rightarrow D^{0}\pi\pi=D^{0}\pi^{+}\pi^{-}+D^{0}\pi^{0}%
\pi^{0}$ & $0.59\pm0.02$ & seen\\\hline
$D_{1}(2420)^{0}\rightarrow D^{+}\pi^{-}\pi^{0}$ & $0.21_{-0.015}^{+0.01}$ &
seen\\\hline
$D_{1}(2420)^{0}\rightarrow D^{+}\pi^{-}$ & $0$ & not seen; $\Gamma(D^{+}%
\pi^{-})/\Gamma(D^{\ast+}\pi^{-})<0.24$\\\hline
$D_{1}(2420)^{+}\rightarrow D^{\ast}\pi=D^{\ast+}\pi^{0}+D^{\ast0}\pi^{+}$ &
$65_{-36}^{+51}$ & $D^{\ast0}\pi^{+}$ seen; full width: $\Gamma=25\pm
6$\\\hline
$D_{1}(2420)^{+}\rightarrow D^{+}\pi\pi=D^{+}\pi^{+}\pi^{-}+D^{+}\pi^{0}%
\pi^{0}$ & $0.56\pm0.02$ & seen\\\hline
$D_{1}(2420)^{+}\rightarrow D^{0}\pi^{0}\pi^{+}$ & $0.22\pm0.01$ &
seen\\\hline
$D_{1}(2420)^{+}\rightarrow D^{0}\pi^{+}$ & $0$ & not seen; $\Gamma(D^{0}%
\pi^{+})/\Gamma(D^{\ast0}\pi^{+})<0.18$\\\hline
$D_{S1}(2536)^{+}\rightarrow D^{\ast}K=D^{\ast0}K^{+}+D^{\ast+}K^{0}$ &
$25_{-15}^{+22}$ & seen; full width $\Gamma=0.92\pm0.03\pm0.04$\\\hline
$D_{S1}(2536)^{+}\rightarrow D^{+}K^{0}$ & $0$ & not seen\\\hline
$\,D_{S1}(2536)^{+}\rightarrow+D^{0}K^{+}$ & $0$ & not seen\\\hline
\end{tabular}

\end{center}

The following comments are in order:

(i) The decay of $D_{0}^{\ast}(2400)^{0}$ into $D\pi$ has a very large
theoretical error due to the imprecise determination of $\phi_{C}$. A
qualitative statement is, however, possible: the decay channel $D_{0}^{\ast
}(2400)^{0}\rightarrow D\pi$ is large and is the only OZI-dominant decay
predicted by our model. This decay channel is also the only one seen in
experiment (although the branching ratio is not yet known). A similar
discussion holds for the charged counterpart $D_{0}^{\ast}(2400)^{+}$.

(ii) The decay widths of the vector charmed states $D^{\ast}(2007)^{0}$ and
$D^{\ast}(2010)^{+}$ are slightly smaller than the experimental results, but
close to the lower bounds of the latter.

(iii) The results for the axial-vector charmed states $D_{1}(2420)^{0}$ and
$D_{1}(2420)^{+}$ are compatible with experiment. Note that the decay into
$D^{\ast}\pi$ is the only one which is experimentally seen. Moreover, the
decays $D_{1}(2420)^{0}\rightarrow D^{+}\pi^{-}$ and $D_{1}(2420)^{+}%
\rightarrow D^{0}\pi^{+}$, although kinematically allowed, do not occur in our
model because there is no respective tree-level coupling; this is in agreement
with the small experimental upper bound. Improvements in the decay channels of
$D_{1}(2420)$ are possible by taking into account also the multiplet of
pseudovector quark-antiquark states. In this way, one will be able to evaluate
the mixing of these configurations and describe at the same time the
resonances $D_{1}(2420)$ and $D_{1}(2430)$.

(iv) In the low-energy language, the vector states $D^{\ast
}(2007)^{0}$ and $D^{\ast}(2010)^{\pm}$ and the axial-vector states
$D_{1}(2420)^{0}$ and $D_{1}(2420)^{+}$ are chiral partners. The fact
that our model is at least in qualitative agreement with the experimental data
for their decay widths indicates that chiral symmetry is still important in the energy
range relevant for charmed mesons.

(v) The decay of the axial-vector strange-charmed $D_{S1}(2536)^{+}\rightarrow
D^{\ast}K$ is too large in our model when compared to the experimental data of
about $1$ MeV. This result is robust upon variation of the parameters, as the
error shows. We thus conclude that the resonance $D_{S1}(2536)^{\pm}$ is not
favored to be (predominantly) a member of the axial-vector multiplet (it can
be, however, a member of the pseudovector multiplet). Then, we discuss two
possible solutions to the problem of identifying the axial-vector
strange-charmed quarkonium:

\textit{Solution 1}: There is a `seed' quark-antiquark axial-vector state
$D_{S1}$ above the $D^{\ast}K$ threshold, which is, however, very broad and
for this reason has \emph{not yet} been detected. Quantum corrections generate
the state $D_{S1}(2460)^{\pm}$ through pole doubling \cite{pennington}. In
this scenario, $D_{S1}(2460)$ is dynamically generated but is still related to
a broad quark-antiquark seed state. In this way, the low mass of
$D_{S1}(2460)$ in comparison to the quark-model prediction \cite{models1} is
due to quantum corrections \cite{loop,coito,lutz}. Then, the state
$D_{S1}(2460)$, being below threshold, has a very small decay width.

\textit{Solution 2}: Also in this case, there is still a broad and not yet
detected quark-antiquark field above threshold, but solution 1 is assumed not
to apply (loops are not sufficient to generate $D_{S1}(2460)$). The resonance
$D_{S1}(2460)^{\pm}$ is not a quark-antiquark field, but a tetraquark or a
loosely bound molecular state and its existence is not related to the
quark-antiquark state of the axial-vector multiplet.

(vi) For the state $D_{S0}^{\ast}(2317)$ similar arguments apply. If the mass
of this state is above the $DK$ threshold, we predict a very large ($\agt 1$
GeV) decay width into $DK$ (for example: $\Gamma_{D_{S0}^{\ast}\rightarrow
DK}\simeq3$ GeV for a $D_{S0}^{\ast}$ mass of $2467$ MeV as determined in
Tab.\ 2). Then, the two solutions mentioned above are applicable also here:

\textit{Solution 1}: A quark-antiquark state with a mass above the $DK$
threshold exists, but it is too broad to be seen in experiment. The state
$D_{S0}(2317)$ arises through the pole-doubling mechanism.

\textit{Solution 2}: Loops are not sufficient to dynamically generate
$D_{S0}^{\ast}(2317).$ The latter is not a quarkonium but either a tetraquark
or a molecular state.

In conclusion, a detailed study of loops in the axial-vector and scalar
strange-charm sector needs to be performed. In the axial-vector strange-charm
sector mixing with a pseudovector quark-antiquark state should also be
included. These tasks go beyond the tree-level analysis of our work but are an
interesting subject for the future. \newline

\section{Conclusion and outlook}

In this work we have developed a four-flavor extended linear sigma model with
vector and axial-vector degrees of freedom. Within this model, we have calculated masses and
decay widths of charmed mesons.

For the coupling constants of the model, 
we have used the values determined in the low-energy study of 
Ref.\ \cite{dick} and listed in Tab.\ 1. The three
remaining parameters related to the current charm quark mass were determined in
a fit to twelve masses of hidden and open charmed mesons. The results are shown in
Tab.\ 2: the open charmed mesons agree within theoretical errors with the
experimental values, while the masses of charmonia are (with the exception of $J/\psi$) underestimated by about 10\%.
The precision of our approach cannot compete with
methods based on heavy-quark symmetry, but it admits a
perspective on charmed states from a low-energy approach based on chiral symmetry 
and dilatation invariance. The level of agreement with experimental data 
proves that these symmetries are, at least to some degree, still
relevant for the charm sector. In this respect,
our approach is a useful tool to investigate the assignment of some charmed
states (see below) and to obtain an independent determination of quantities
such as the chiral condensate of charm-anticharm quarks. The latter turns out
to be sizable, showing that the charm quark, although heavy, is indeed still
connected to nontrivial vacuum dynamics.

We have also calculated the weak-decay constants of the
pseudoscalar states $D,$ $D_{S},$ and $\eta_{c}$, which are in fair agreement
with the experimental values and, as a last step, we have evaluated the
OZI-dominant decays of charmed mesons (Tab.\ 3). The result for $D_{0}^{\ast
}(2400)^{0},$ $D_{0}^{\ast}(2400)^{+},$ $D^{\ast}(2007)^{0},$ $D^{\ast
}(2010)^{+},$ $D_{1}(2420)^{0},$ and $D_{1}(2420)^{+}$ are compatible with the
results listed by the PDG \cite{PDG}, although the
theoretical errors are still quite large and some experimental values are only upper bounds.

In summary, the fact that an (although at this stage only rough) qualitative
description is obtained by using a chiral model and, more remarkably, employing
values for the coupling constants, which were determined in a study of 
$N_{f}=3$ mesons, means that a remnant
of chiral symmetry is present also in the sector of charmed mesons. Chiral
symmetry is still of some relevance because the parameters of the eLSM
do not vary too much as a function of the energy at which they are probed.
Besides mass terms which describe the large contribution of the current charm
quark mass, all interaction terms are the same as in the low-energy effective
model of Refs.\ \cite{denis,stani,dick} which was built under the requirements
of chiral symmetry and dilatation invariance. As a by-product of our work we
also evaluate the charm condensate which is of the same order as the
nonstrange and strange quark condensates. This is also in accord with chiral
dynamics enlarged to the group $U(4)_{R}\times U(4)_{L}$.

Concerning the assignment of the scalar and axial-vector strange-charmed
quarkonium states $D_{S0}$ and $D_{S1},$ we obtain the following: If the
masses of these quarkonia are above the respective thresholds, we find that
their decay widths are too large, which probably means that these states, even
if they exist, have escaped detection. In this case, the resonances
$D_{S0}^{\ast}(2317)$ and $D_{S1}(2460)$ can emerge as dynamically generated
companion poles (alternatively, they can be tetraquark or molecular states).
Our results imply also that the interpretation of the resonance $D_{S1}(2536)$
as a member of the axial-vector multiplet is not favored because the
experimental width is too narrow when compared to the theoretical width of a
quarkonium state with the same mass. An investigation of these resonances
necessitates the calculation of quantum fluctuations and represents a topic of
future work.

A further important future project is the study of OZI-suppressed decays of
charmonium states. To this end, one should allow for a nonzero value of the
large-$N_{c}$ suppressed parameters. Due to the large amount of existing data,
this is an interesting project to test our chiral approach in more detail in
the realm of hidden and open charmed states.

\section*{Acknowledgments}

The authors thank F.-K.\ Guo for pointing us to Ref.\ \cite{fkguo}, 
and C.\ Sasaki and J.\ Schaffner-Bielich for making us aware of Refs.\
\cite{mishra,gottfriedklevansky}. We also acknowledge useful discussions with
D.\ Parganlija, S.\ Schramm, and M.\ Wagner about the
extension of the eLSM to four flavors, and discussions with P.\ Kovacs and Gy.\ Wolf
about the fit in the three-flavor case. W.I.E.\ acknowledges
financial support from DAAD and HGS-HIRe.

\appendix

\section{Shifts and masses}

\subsection{Shifts}

Due to spontaneous symmetry breaking the scalar-isoscalar fields $\sigma_{N}$,
$\sigma_{S}$, and $\chi_{c0}$ are shifted by their vacuum expectation values
$\phi_{N}$, $\phi_{S}$, and $\phi_{C}$ as:%
\begin{equation}
\sigma_{N}\rightarrow\sigma_{N}+\phi_{N}\,\,\,\text{,}\,\,\,\sigma
_{S}\rightarrow\sigma_{S}+\phi_{S},\text{ }\chi_{c0}\rightarrow\chi_{c0}%
+\phi_{C}\,. \label{9}%
\end{equation}
As a consequence, bilinear mixing terms involving the mesons $\eta_{N}%
$-$f_{1N}$, $\overrightarrow{\pi}$-$\overrightarrow{a}_{1}$, $\eta_{S}%
$-$f_{1S}$, $K^{\ast}_{0}$-$K^{\ast}$, and $K$-$K_{1}$ arise
\cite{denis,stani,dick,three flavor,detail mixing}:
\begin{align}
&  -g_{1}\phi_{N}(f_{1N}^{\mu}\partial_{\mu}\eta_{N}+\overrightarrow{a}%
_{1}^{\mu}\cdot\partial_{\mu}\overrightarrow{\pi}) -\sqrt{2}\,g_{1}\phi
_{S}f^{\mu}_{1S}\partial_{\mu}\eta_{S}\nonumber\\
&  -i\frac{g_{1}}{2}(\phi_{N}-\phi_{S})(\overline{K}\,^{*\mu0}\,\partial_{\mu
}K^{\ast0}_{0}+K^{*\mu-}\,\partial_{\mu}K^{\ast+}_{0})\nonumber\\
&  +i\frac{g_{1}}{2}(\phi_{N}-\sqrt{2}\phi_{s})(K^{*\mu0}\,\partial_{\mu
}\overline{K}^{\ast0}_{0}+K^{*\mu+}\, \partial_{\mu}K^{\ast-}_{0})\nonumber\\
&  -\frac{g_{1}}{2}(\phi_{N}+\sqrt{2}\,\phi_{S})(K_{1}^{\mu0}\,\partial_{\mu
}\overline{K}^{0}+K_{1}^{\mu+}\,\partial_{\mu}K^{-}+\overline{K}_{1}^{\mu
0}\,\partial_{\mu}K^{0}+K_{1}^{\mu-}\,\partial_{\mu}K^{+})\,. \label{mixing1}%
\end{align}
In addition, for charmed mesons similar mixing terms of the type $\eta_{C}%
$-$\chi_{c1}$, $D_{S}$-$D_{S1}$, $D_{S0}^{\ast}$-$D_{S1}^{\ast}$, $D_{0}%
^{\ast}$-$D^{\ast}$, and $D$-$D_{1}$ are present:%
\begin{align}
&  -g_{1}\phi_{C}\,\chi_{c1}^{\mu}\,\partial_{\mu}\eta_{C}-\frac{g_{1}}%
{\sqrt{2}}\,g_{1} \phi_{S}(D_{S1}^{\mu-}\,\partial_{\mu}D_{S}^{+}+D_{S1}%
^{\mu+}\,\partial_{\mu}D_{S}^{-})\nonumber\\
&  +i\frac{g_{1}}{\sqrt{2}}\,\phi_{S}(D_{S}^{\ast\mu-}\,\partial_{\mu}%
D_{S0}^{\ast+}-D_{S}^{\ast\mu+}\,\partial_{\mu}D_{S0}^{\ast-})\nonumber\\
&  +\,i\frac{g_{1}}{2}\,\phi_{N}(D^{*\mu-}\,\partial_{\mu}D^{*+}_{0}-D^{*\mu
+}\, \partial_{\mu}D^{*-}_{0}+ D^{*\mu0}\,\partial_{\mu}\overline{D}%
\,^{*0}_{0}-\overline{D}\,^{*\mu0}\,\partial_{\mu}D^{*0}_{0})\nonumber\\
&  -\frac{g_{1}}{2}\,\phi_{N}(D^{0\mu}_{1}\, \partial_{\mu}\overline{D}%
^{0}+\overline{D}\,^{\mu0}_{1}\, \partial_{\mu}D^{0}+D^{\mu+}_{1}\,
\partial_{\mu}D^{-}+D^{\mu-}_{1}\, \partial_{\mu}D^{+})\text{ .}%
\end{align}

These mixing terms are removed by performing the following field transformations:%

\begin{align}
&  f_{1N,S}^{\mu}\rightarrow f_{1N,S}^{\mu}+w_{f_{1N,S}}\,Z_{\eta_{N,S}%
}\,\partial^{\mu}\eta_{N,S}\,,\\
&  \overrightarrow{a}_{1}^{\mu}\rightarrow\overrightarrow{a} _{1}^{\mu
}+w_{a_{1}}\,Z_{\pi}\,\partial^{\mu}\overrightarrow{\pi}\,,\\
&  K^{*\mu+,0}\rightarrow K^{*\mu+,0}+w_{K^{*}}\,Z_{K^{\ast}_{0}}%
\,\partial^{\mu}K^{\ast+,0}_{0}\,,\\
&  K^{*\mu-,\bar{0}}\rightarrow K^{*\mu-,\bar{0}}+w^{\ast}_{K^{*}}%
\,Z_{K^{\ast}_{0}}\,\partial^{\mu}K^{\ast-,\bar{0}}_{0}\,,\\
&  K^{\mu\pm,0,\bar{0}}_{1}\rightarrow K^{\mu\pm,0,\bar{0}}_{1}+w_{K_{1}%
}\,Z_{K}\,\partial^{\mu}K^{\pm,0,\bar{0}}\,,\\
&  \chi_{c1}^{\mu}\rightarrow\chi_{c1}^{\mu}+w_{\chi_{c1}}\,Z_{\eta_{C}}
\,\partial^{\mu}\eta_{C}\,,\\
&  D_{S1}^{\mu\pm}\rightarrow D_{S1}^{\mu\pm}+w_{D_{S1}}\,Z_{D_{S}}
\,\partial^{\mu}D_{S}^{\pm}\>,\\
&  D^{*\mu-}_{S}\rightarrow D^{*\mu-}_{S}+w_{D^{*}_{S}}\,Z_{D^{*}_{S0}}
\,\partial^{\mu}D^{*-}_{S0}\,,\\
&  D^{*\mu+}_{S}\rightarrow D^{*\mu+}_{S}+w^{*}_{D^{*}_{S}}\,Z_{D^{*}_{S0}}
\,\partial^{\mu}D^{*+}_{S0}\>,\\
&  D^{*\mu+}\rightarrow D^{*\mu+}+w^{*}_{D^{*}}\,Z_{D^{*}_{0}} \,\partial
^{\mu}D^{*+}_{0}\,,\\
&  D^{*\mu-}\rightarrow D^{*\mu-}+w_{D^{*}}\,Z_{D^{*}_{0}} \,\partial^{\mu
}D^{*-}_{0}\>,\\
&  \overline{D}\,^{*\mu0}\rightarrow\overline{D}\,^{*\mu0}+w^{*}_{D^{*0}%
}\,Z_{D^{*0}_{0}} \,\partial^{\mu}\overline{D}\,^{*0}_{0}\,,\\
&  D^{*\mu0}\rightarrow D^{*\mu0}+w_{D^{*0}}\,Z_{D^{*0}_{0}} \,\partial^{\mu
}D^{*0}_{0}\>,\\
&  D_{1}^{\mu\pm,0,\bar{0}}\rightarrow D_{1}^{\mu\pm,0,\bar{0}}+w_{D_{1}%
}\,Z_{D} \,\partial^{\mu}D^{\pm,0,\bar{0}}\>,
\end{align}

The constants appearing in the previous expressions are:
\begin{align}
&  w_{f_{1N}}=w_{a_{1}}=\frac{g_{1}\phi_{N}}{m_{a_{1}}^{2}}%
,\,\,\,\,\,\,\,\,\,\,\,\,\,\,\,\,\,\,\,\,\,\,\,\,\,\,w_{f_{1S}}=\frac{\sqrt
{2}g_{1}\phi_{S}}{m_{f_{1S}}^{2}}\>,\\
&  w_{K^{\ast}}=\frac{ig_{1}(\phi_{N}-\sqrt{2}\phi_{S})}{2\,m_{K^{\ast}}^{2}%
},\,\,\,\,\,\,\,\,\,\,\,\,\,\,\,\,w_{K_{1}}=\frac{g_{1}(\phi_{N}+\sqrt{2}%
\phi_{S}}{2\,m_{K_{1}}^{2}}\>,\\
&  w_{\chi_{c1}}=\frac{\sqrt{2}g_{1}\phi_{C}}{m_{\chi_{c1}}^{2}}%
,\,\,\,\,\,\,\,\,\,\,\,\,\,\,\,\,\,\,\,\,\,\,\,\,\,\,\,\,\,\,\,\,\,\,\,w_{D_{S1}%
}=\frac{g_{1}(\phi_{S}+\phi_{C})}{\sqrt{2}m_{D_{S1}}^{2}}\>,\\
&  w_{D_{S}^{\ast}}=\frac{ig_{1}(\phi_{S}-\phi_{C})}{\sqrt{2}m_{D_{S}^{\ast}%
}^{2}},\,\,\,\,\,\,\,\,\,\,\,\,\,\,\,\,\,\,\,\,\,\,\,w_{D^{\ast}}=\frac
{ig_{1}(\phi_{N}-\sqrt{2}\phi_{C})}{2\,m_{D^{\ast}}^{2}}\>,\\
&  w_{D^{\ast0}}=\frac{ig_{1}(\phi_{N}-\sqrt{2}\phi_{C})}{2\,m_{D^{\ast0}}%
^{2}},\,\,\,\,\,\,\,\,\,\,\,\,\,\,w_{D_{1}}=\frac{g_{1}(\phi_{N}+\sqrt{2}%
\phi_{C})}{2\,m_{D_{1}}^{2}}\>.
\end{align}
Moreover, one has to rescale the (pseudo)scalar fields as:
\begin{align}
&  \pi^{\pm,0}\rightarrow Z_{\pi}\pi^{\pm,0}%
,\,\,\,\,\,\,\,\,\,\,\,\,\,\,\,\,\,\,\,\,\,\,\,\,\,\,\,\,\,\,\,\,\,\,\,\,\,\,\,\,\,\,\,\,K^{\pm
,0,\bar{0}}\rightarrow Z_{K}K^{\pm,0,\bar{0}}\text{ },\nonumber\\
&  \eta_{N/S/C}\rightarrow Z_{\eta_{N} /\eta_{S}/\eta_{C}}\eta_{N/S/C}%
,\,\,\,\,\,\,\,\,\,\,{K^{\star}_{0}}^{\pm,0,\bar{0}}\rightarrow Z_{K^{\star}%
}{K^{\star}_{0}}^{\pm,0,\bar{0}}\,,\\
&  D^{\pm,0,\bar{0}}\rightarrow Z_{D}D^{\pm,0,\bar{0}}%
,\,\,\,\,\,\,\,\,\,\,\,\,\,\,\,\,\,\,\,\,\,\,\,\,\,\,\,\,\,\,\,\,D^{*\pm}%
_{0}\rightarrow Z_{D^{*}_{0}}D^{*\pm}_{0}\>,\nonumber\\
&  D_{0}^{\ast0,\bar{0}}\rightarrow Z_{D_{0}^{\ast0}}D_{0}^{\ast0,\bar{0}%
},\,\,\,\,\,\,\,\,\,\,\,\,\,\,\,\,\,\,\,\,\,\,\,\,\,\,\,\,\,\,\,\,D_{S0}%
^{\ast\pm}\rightarrow Z_{D_{S0}^{\ast}}D_{S0}^{\ast\pm}\text{ ,}%
\end{align}
where the wave-function renormalization constants read:%

\begin{align}
&  Z_{\pi}\equiv Z_{\eta_{N}}=\frac{m_{a_{1}}}{\sqrt{m_{a_{1}}^{2}-g_{1}%
^{2}\,\phi_{N}^{2}}}%
,\,\,\,\,\,\,\,\,\,\,\,\,\,\,\,\,\,\,\,\,\,\,\,\,\,\,\,\,\,\,\,\,Z_{\eta_{S}%
}=\frac{m_{f_{1S}}}{\sqrt{m_{f_{1S}}^{2}-2\,g_{1}^{2}\,\phi_{S}^{2}}}\>,\\
&  Z_{K}=\frac{2m_{K_{1}}}{\sqrt{4m_{K_{1}}^{2}-g_{1}^{2}(\phi_{N}+\sqrt
{2}\phi_{S})^{2}}},\,\,\,\,\,\,\,\,\,\,\,\,\,\,\,\,Z_{K_{S}}=\frac
{2m_{K_{\ast}}}{\sqrt{4m_{K_{\ast}}^{2}-g_{1}^{2}(\phi_{N}-\sqrt{2}\phi
_{S})^{2}}}\>,\\
&  Z_{\eta_{C}}=\frac{m_{\chi_{c1}}}{\sqrt{m_{\chi_{c1}}^{2}-2g_{1}^{2}%
\phi_{C}^{2}}}%
\,,\,\,\,\,\,\,\,\,\,\,\,\,\,\,\,\,\,\,\,\,\,\,\,\,\,\,\,\,\,\,\,\,\,\,\,\,\,\,\,\,\,\,Z_{D_{S}%
}=\frac{\sqrt{2}m_{D_{S1}}}{\sqrt{2m_{D_{S1}}^{2}-g_{1}^{2}(\phi_{S}+\phi
_{C})^{2}}}\>,\\
&  Z_{D_{S0}^{\ast}}=\frac{\sqrt{2}m_{D_{S}^{\ast}}}{\sqrt{2m_{D_{S}^{\ast}%
}^{2}-g_{1}^{2}(\phi_{S}-\phi_{C})^{2}}}\text{ ,}%
\,\,\,\,\,\,\,\,\,\,\,\,\,\,\,\,\,\,Z_{D_{0}^{\ast}}=\frac{2\,m_{D^{\ast}}%
}{\sqrt{4m_{D^{\ast}}^{2}-g_{1}^{2}(\phi_{N}-\sqrt{2}\phi_{C})^{2}}}\>,\\
&  Z_{D_{0}^{\ast0}}=\frac{2\,m_{D^{\ast0}}}{\sqrt{4m_{D^{\ast0}}^{2}%
-g_{1}^{2}(\phi_{N}-\sqrt{2}\phi_{C})^{2}}}\,,\,\,\,\,\,\,\,\,\,Z_{D}%
=\frac{2\,m_{D_{1}}}{\sqrt{4m_{D_{1}}^{2}-g_{1}^{2}(\phi_{N}+\sqrt{2}\phi
_{C})^{2}}}\,.
\end{align}

The nonstrange, strange, and charm condensates read:%
\begin{equation}
\phi_{N}=Z_{\pi}f_{\pi}\text{ },
\end{equation}
\begin{equation}
\phi_{S}=\frac{2\,Z_{K}f_{K}-\phi_{N}}{\sqrt{2}}\text{ },
\end{equation}

\begin{equation}
\phi_{C}=\frac{2Z_{D}f_{D}-\phi_{N}}{\sqrt{2}}=\sqrt{2}Z_{D_{S}}f_{D_{S}}%
-\phi_{S}=\frac{Z_{\eta_{C}}f_{\eta_{C}}}{\sqrt{2}},
\end{equation}
The quantities $f_{\pi}=92.4$ MeV and $f_{K}=155/\sqrt{2}$ MeV are the pion
and kaon decay constants, while $f_{D}$ and $f_{D_{S}}$ are the decay
constants of the pseudoscalar $D$ and $D_{S}$ mesons, respectively.

\subsection{Tree-level masses}

After having performed the transformation above, we obtain the tree-level
masses of nonstrange-strange mesons in the eLSM:

(i) Pseudoscalar mesons:
\begin{align}
m_{\pi}^{2}  &  =Z_{\pi}^{2}\bigg[m_{0}^{2}+\bigg(\lambda_{1}+\frac
{\lambda_{2}}{2}\bigg)\,\phi_{N}^{2}+\lambda_{1}\,\phi_{S}^{2}+\lambda
_{1}\,\phi_{C}^{2}\bigg]\>,\label{45}\\
m_{K}^{2}  &  =Z_{K}^{2}\bigg[m_{0}^{2}+\bigg(\lambda_{1}+\frac{\lambda_{2}%
}{2}\bigg)\,\phi_{N}^{2}-\frac{\lambda_{2}}{\sqrt{2}}\,\phi_{N}\,\phi
_{S}+(\lambda_{1}+\lambda_{2})\,\phi_{S}^{2}+\lambda_{1}\,\phi_{C}%
^{2}\bigg]\>,\label{46}\\
m_{\eta_{N}}^{2}  &  =Z_{\pi}^{2}\bigg[m_{0}^{2}+\bigg(\lambda_{1}%
+\frac{\lambda_{2}}{2}\bigg)\,\phi_{N}^{2}+\lambda_{1}\,\phi_{S}^{2}%
+\lambda_{1}\,\phi_{C}^{2}+\frac{c}{2}\,\phi_{N}^{2}\,\phi_{S}^{2}\,\phi
_{C}^{2}\bigg]\>,\label{47}\\
m_{\eta_{S}}^{2}  &  =Z_{\eta_{S}}^{2}\bigg[m_{0}^{2}+\lambda_{1}\,\phi
_{N}^{2}+(\lambda_{1}+\lambda_{2})\,\phi_{S}^{2}+\lambda_{1}\,\phi_{C}%
^{2}+\frac{c}{8}\,\phi_{N}^{4}\,\phi_{C}^{2}\bigg]\>. \label{48}%
\end{align}

(ii) Scalar mesons:
\begin{align}
m_{a_{0}}^{2}  &  =m_{0}^{2}+\bigg(\lambda_{1}+\frac{3}{2}\lambda
_{2}\bigg)\phi_{N}^{2}+\lambda_{1}\,\phi_{S}^{2}+\lambda_{1}\,\phi_{C}%
^{2}\>,\label{41}\\
m_{K_{0}^{\ast}}^{2}  &  =Z_{K_{0}^{\ast}}^{2}\bigg[m_{0}^{2}+\bigg(\lambda
_{1}+\frac{\lambda_{2}}{2}\bigg)\,\phi_{N}^{2}+\frac{\lambda_{2}}{\sqrt{2}%
}\,\phi_{N}\,\phi_{S}+(\lambda_{1}+\lambda_{2})\,\phi_{S}^{2}+\lambda
_{1}\,\phi_{C}^{2}\bigg]\>,\label{42}\\
m_{\sigma_{N}}^{2}  &  =m_{0}^{2}+3\bigg(\lambda_{1}+\frac{\lambda_{2}}%
{2}\bigg)\,\phi_{N}^{2}+\lambda_{1}\,\phi_{S}^{2}+\lambda_{1}\,\phi_{C}%
^{2}\>,\label{43}\\
m_{\sigma_{S}}^{2}  &  =m_{0}^{2}+\lambda_{1}\phi_{N}^{2}+3(\lambda
_{1}+\lambda_{2})\,\phi_{S}^{2}+\lambda_{1}\,\phi_{C}^{2}\>\text{ .}
\label{44}%
\end{align}

(iii) Vector mesons:%
\begin{align}
m_{\rho}^{2}  &  =m_{\omega_{N}}^{2}\;,\\
m_{\omega_{N}}^{2}  &  =m_{1}^{2}+2\,\delta_{N}+\frac{\phi_{N}^{2}}{2}%
\,(h_{1}+h_{2}+h_{3})+\frac{h_{1}}{2}\,\phi_{S}^{2}+\frac{h_{1}}{2}\,\phi
_{C}^{2}\;,\label{m_rho}\\
m_{\omega_{S}}^{2}  &  =m_{1}^{2}+2\,\delta_{S}+\frac{h_{1}}{2}\,\phi_{N}%
^{2}+\bigg(\frac{h_{1}}{2}+h_{2}+h_{3}\bigg)\,\phi_{S}^{2}+\frac{h_{1}}%
{2}\,\phi_{C}^{2}\>,\label{34}\\
m_{K^{\ast}}^{2}  &  =m_{1}^{2}+\delta_{N}+\delta_{S}+\frac{\phi_{N}^{2}}%
{2}\bigg(\frac{g_{1}^{2}}{2}+h_{1}+\frac{h_{2}}{2}\bigg)+\frac{1}{\sqrt{2}%
}\,\phi_{N}\,\phi_{S}\,(h_{3}-g_{1}^{2})+\frac{\phi_{S}^{2}}{2}(g_{1}%
^{2}+h_{1}+h_{2})+\frac{h_{1}}{2}\,\phi_{C}^{2}\>\text{ .} \label{36}%
\end{align}

(iv) Axial-vector mesons:
\begin{align}
m_{f_{1N}}^{2}  &  =m_{a_{1}}^{2},\label{38}\\
m_{a_{1}}^{2}  &  =m_{1}^{2}+2\delta_{N}+g_{1}^{2}\phi_{N}^{2}+\frac{\phi
_{N}^{2}}{2}(h_{1}+h_{2}-h_{3})+\frac{h_{1}}{2}\,\phi_{S}^{2}+\frac{h_{1}}%
{2}\,\phi_{C}^{2}\>,\label{37}\\
m_{f_{1S}}^{2}  &  =m_{1}^{2}+2\delta_{S}+\frac{h_{1}}{2}\,\phi_{N}^{2}%
+\frac{h_{1}}{2}\,\phi_{C}^{2}+2g_{1}^{2}\,\phi_{S}^{2}+\phi_{S}%
^{2}\bigg(\frac{h_{1}}{2}+\,h_{2}-h_{3}\bigg)\>,\label{39}\\
m_{K_{1}}^{2}  &  =m_{1}^{2}+\delta_{N}+\delta_{S}+\frac{\phi_{N}^{2}}%
{2}\bigg(\frac{g_{1}^{2}}{2}+h_{1}+\frac{h_{2}}{2}\bigg)+\frac{1}{\sqrt{2}%
}\,\phi_{N}\,\phi_{S}\,(g_{1}^{2}-h_{3})+\frac{\phi_{S}^{2}}{2}(g_{1}%
^{2}+h_{1}+h_{2})+\frac{h_{1}}{2}\,\phi_{C}^{2}\>, \label{40}%
\end{align}

All previous expressions coincide with Ref.\ \cite{dick} for $h_1 =0$.

The masses of (open and hidden) charmed mesons are as follows.

(i) Pseudoscalar charmed mesons:
\begin{align}
m_{\eta_{C}}^{2}  &  =Z_{\eta_{C}}^{2}[m_{0}^{2}+\lambda_{1}\phi_{N}%
^{2}+\lambda_{1}\phi_{S}^{2}+(\lambda_{1}+\lambda_{2})\phi_{C}^{2}+\frac{c}%
{8}\,\phi_{N}^{4}\,\phi_{S}^{2}+2\,\varepsilon_{C}]\>,\label{61}\\
m_{D}^{2}  &  =Z_{D}^{2}\bigg[m_{0}^{2}+\bigg(\lambda_{1}+\frac{\lambda_{2}%
}{2}\bigg)\phi_{N}^{2}+\lambda_{1}\phi_{S}^{2}-\frac{\lambda_{2}}{\sqrt{2}%
}\phi_{N}\phi_{C}+(\lambda_{1}+\lambda_{2})\phi_{C}^{2}+\varepsilon
_{C}\bigg]\>,\label{62}\\
m_{D_{S}}^{2}  &  =Z_{D_{S}}^{2}[m_{0}^{2}+\lambda_{1}\phi_{N}^{2}%
+(\lambda_{1}+\lambda_{2})\phi_{S}^{2}-\lambda_{2}\phi_{C}\phi_{S}%
+(\lambda_{1}+\lambda_{2})\phi_{C}^{2}+\varepsilon_{C}]\>\text{ .} \label{63}%
\end{align}

(ii) Scalar charmed mesons:%
\begin{align}
m_{\chi_{c0}}^{2}  &  =m_{0}^{2}+\lambda_{1}\phi_{N}^{2}+\lambda_{1}\phi
_{S}^{2}+3(\lambda_{1}+\lambda_{2})\phi_{C}^{2}+2\,\varepsilon_{C}\text{
},\label{57}\\
m_{D_{0}^{\ast}}^{2}  &  =Z_{D_{0}^{\ast}}^{2}\bigg[m_{0}^{2}+\bigg(\lambda
_{1}+\frac{\lambda_{2}}{2}\bigg)\phi_{N}^{2}+\lambda_{1}\phi_{S}^{2}%
+\frac{\lambda_{2}}{\sqrt{2}}\phi_{C}\phi_{N}+(\lambda_{1}+\lambda_{2}%
)\phi_{C}^{2}+\varepsilon_{C}\bigg]\>,\label{59}\\
m_{D_{S0}^{\ast}}^{2}  &  =Z_{D_{S0}}^{2}[m_{0}^{2}+\lambda_{1}\phi_{N}%
^{2}+(\lambda_{1}+\lambda_{2})\phi_{S}^{2}+\lambda_{2}\phi_{C}\phi
_{S}+(\lambda_{1}+\lambda_{2})\phi_{C}^{2}+\varepsilon_{C}]\>\text{ .}
\label{60}%
\end{align}

(iii) Vector charmed mesons:
\begin{align}
m_{D^{\ast}}^{2}  &  =m_{1}^{2}+\delta_{N}+\delta_{C}+\frac{\phi_{N}^{2}}%
{2}\bigg(\frac{g_{1}^{2}}{2}+h_{1}+\frac{h_{2}}{2}\bigg)+\frac{1}{\sqrt{2}%
}\,\phi_{N}\,\phi_{C}(h_{3}-\,g_{1}^{2})+\frac{\phi_{C}^{2}}{2}(g_{1}%
^{2}+h_{1}+h_{2})+\frac{h_{1}}{2}\,\phi_{S}^{2}\>,\label{51}\\
m_{J/\psi}^{2}  &  =m_{1}^{2}+2\delta_{C}+\frac{h_{1}}{2}\,\phi_{N}^{2}%
+\frac{h_{1}}{2}\,\phi_{S}^{2}+\bigg(\frac{h_{1}}{2}+h_{2}+h_{3}%
\bigg)\,\phi_{C}^{2}\>,\label{52}\\
m_{D_{S}^{\ast}}^{2}  &  =m_{1}^{2}+\delta_{S}+\delta_{C}+\frac{\phi_{S}^{2}%
}{2}(g_{1}^{2}+h_{1}+h_{2})+\phi_{S}\,\phi_{C}(h_{3}-g_{1}^{2})+\frac{\phi
_{C}^{2}}{2}(g_{1}^{2}+h_{1}+h_{2})+\frac{h_{1}}{2}\,\phi_{N}^{2}\>\text{ .}
\label{53}%
\end{align}

(iv) Axial-vector charmed mesons:
\begin{align}
m_{D_{S1}}^{2}  &  =m_{1}^{2}+\delta_{S}+\delta_{C}+\frac{\phi_{S}^{2}}%
{2}(g_{1}^{2}+h_{1}+h_{2})+\phi_{S}\,\phi_{C}(g_{1}^{2}-h_{3})+\frac{\phi
_{C}^{2}}{2}(g_{1}^{2}+h_{1}+h_{2})+\frac{h_{1}}{2}\,\phi_{N}^{2}%
\>,\label{54}\\
m_{D_{1}}^{2}  &  =m_{1}^{2}+\delta_{N}+\delta_{C}+\frac{\phi_{N}^{2}}%
{2}\bigg(\frac{g_{1}^{2}}{2}+h_{1}+\frac{h_{2}}{2}\bigg)+\frac{1}{\sqrt{2}%
}\,\phi_{N}\,\phi_{C}(g_{1}^{2}-h_{3})+\frac{\phi_{C}^{2}}{2}(g_{1}^{2}%
+h_{1}+h_{2})+\frac{h_{1}}{2}\,\phi_{S}^{2}\>,\label{55}\\
m_{\chi_{c1}}^{2}  &  =m_{1}^{2}+2\delta_{C}+\frac{h_{1}}{2}\,\phi_{N}%
^{2}+\frac{h_{1}}{2}\phi_{S}^{2}+2g_{1}^{2}\,\phi_{C}^{2}+\phi_{C}%
^{2}\bigg(\frac{h_{1}}{2}+h_{2}-h_{3}\bigg)\>\text{ .} \label{56}%
\end{align}

\section{Decay widths}

\subsection{General formula}

The decay width of a particle $A$ decaying into particles $B$ and $C$ has the
general expression%

\begin{equation}
\Gamma_{A\rightarrow BC}=\frac{S_{A\rightarrow BC}K(m_{A},\,m_{B},\,m_{C}%
)}{8\pi m_{A}^{2}}|\mathcal{M}_{A\rightarrow BC}|^{2}\>, \label{B1}%
\end{equation}
where $K(m_{A},\,m_{B},\,m_{C})$ is the modulus of the three-momentum of one
of the outgoing particles:
\begin{equation}
K(m_{A},\,m_{B},\,m_{C})=\frac{1}{2m_{A}}\sqrt{m_{A}^{4}+(m_{B}^{2}-m_{C}%
^{2})^{2}-2m_{A}^{2}\,(m_{B}^{2}+m_{C}^{2})}\theta(m_{A}-m_{B}-m_{C})\text{ ,}
\label{B2}%
\end{equation}
and where $\mathcal{M}_{A\rightarrow BC}$ is the corresponding decay
amplitude. The quantity $S_{A\rightarrow BC}$ refers to a symmetrization
factor (it equals $1$ if $B$ and $C$ are different and $2$ if they are identical).

The decay width of $A$ into three particles $B_{1},$ $B_{2},$ and $B_{3}$
takes the general expression \cite{PDG}%
\begin{equation}
\Gamma_{A\rightarrow B_{1}B_{2}B_{3}}=\frac{S_{A\rightarrow B_{1}B_{2}B_{3}}%
}{32(2\pi)^{3}m_{A}^{3}}\int_{(m_{1}+m_{2})^{2}}^{(m_{A}-m_{3})^{2}}%
\int_{(m_{23})_{\min}}^{(m_{23})_{\max}}|-i\mathcal{M}_{A\rightarrow
B_{1}B_{2}B_{3}}|^{2}\,dm_{23}^{2}\,dm_{12}^{2}\,, \label{B3}%
\end{equation}
where
\begin{align}
(m_{23})_{\min}  &  =(E_{2}^{\ast}+E_{3}^{\ast})^{2}-\left(  \sqrt{E_{2}%
^{\ast2}-m_{2}^{2}}+\sqrt{E_{3}^{\ast2}-m_{3}^{2}}\right)  ^{2}\text{ ,}\\
(m_{23})_{\max}  &  =(E_{2}^{\ast}+E_{3}^{\ast})^{2}-\left(  \sqrt{E_{2}%
^{\ast2}-m_{2}^{2}}-\sqrt{E_{3}^{\ast2}-m_{3}^{2}}\right)  ^{2}\text{ ,}%
\end{align}
and%
\begin{equation}
E_{2}^{\ast}=\frac{m_{12}^{2}-m_{1}^{2}+m_{2}^{2}}{2m_{12}}\text{ , }%
E_{3}^{\ast}=\frac{m_{A}^{2}-m_{12}^{2}-m_{3}^{2}}{2m_{12}}\text{ .}
\label{B6}%
\end{equation}
The quantities $m_{1},$ $m_{2},$ $m_{3}$ are the masses of $B_{1},$ $B_{2}$,
and $B_{3}$, $\mathcal{M}_{A\rightarrow B_{1}B_{2}B_{3}}$ is the tree-level
decay amplitude, and $S_{A\rightarrow B_{1}B_{2}B_{3}}$ is a symmetrization
factor (it equals $1$ if $B_{1},B_{2}$, and $B_{3}$ are all different, it
equals $2$ for two identical particles in the final state, and it equals $6$
for three identical particles in the final state).

\subsection{Decay widths of charmed scalar mesons}

\label{app4} In the scalar sector we report the expressions for the decay
widths of $D_{0}^{\ast0},\,D_{0}^{\ast+}$, and $D_{S0}^{\ast+}$.

The scalar state $D_{0}^{\ast0}$ decays into $D^{0}\pi^{0}$ and$\,D^{+}\pi
^{-}$. The explicit expression for the process $D_{0}^{\ast0}\rightarrow
D^{0}\pi^{0}$ reads\newline%
\begin{equation}%
\begin{split}
\Gamma_{S\rightarrow P_{1}P_{2}}=\frac{1}{8\pi m_{S}}\left[  \frac{(m_{S}%
^{2}-m_{P_{1}}^{2}-m_{P_{2}}^{2})^{2}-4m_{P_{2}}^{2}m_{P_{1}}^{2}}{4m_{S}^{4}%
}\right]  ^{1/2}  &  \left[  A_{D_{0}^{\ast0}D\pi}+(C_{D_{0}^{\ast0}D\pi
}+E_{D_{0}^{\ast0}D\pi}-B_{D_{0}^{\ast0}D\pi})\right. \\
&  \times\left.  \frac{m_{S}^{2}-m_{P_{1}}^{2}-m_{P_{2}}^{2}}{2}%
+C_{D_{0}^{\ast0}D\pi}m_{P_{1}}^{2}+E_{D_{0}^{\ast0}D\pi}m_{P_{2}}^{2}\right]
^{2}\;,
\end{split}
\label{d0star}%
\end{equation}
where $S,\,P_{1},\,$and $P_{2}$ refer to the scalar meson $D_{0}^{\ast0}$ and
to the pseudoscalar mesons $D^{0}$ and $\pi^{0}$ and where%

\begin{align}
A_{D_{0}^{\ast0}D\pi}  &  =-\frac{Z_{\pi}Z_{D}Z_{D_{0}^{\ast0}}}{\sqrt{2}%
}\,\lambda_{2}\phi_{C},\\
B_{D_{0}^{\ast0}D\pi}  &  =\frac{Z_{\pi}Z_{D}Z_{D_{0}^{\ast0}}}{4}\,w_{a_{1}%
}w_{D_{1}}\left[  g_{1}^{2}(3\phi_{N}+\sqrt{2}\phi_{C})-2g_{1}\frac{w_{a_{1}%
}+w_{D_{1}}}{w_{a_{1}}w_{D_{1}}}+h_{2}(\phi_{N}+\sqrt{2}\phi_{C})-2h_{3}%
\phi_{N}\right]  ,\\
C_{D_{0}^{\ast0}D\pi}  &  =-\frac{Z_{\pi}Z_{D}Z_{D_{0}^{\ast0}}}{2}%
w_{D^{\star0}}w_{D_{1}}\left[  \sqrt{2}ig_{1}^{2}\phi_{C}-g_{1}\frac{w_{D_{1}%
}+iw_{D^{\star0}}}{w_{D^{\star0}}w_{D_{1}}}-\sqrt{2}ih_{3}\phi_{C}\right]  ,\\
E_{D_{0}^{\ast0}D\pi}  &  =\frac{Z_{\pi}Z_{D}Z_{D_{0}^{\ast0}}}{4}%
w_{D^{\star0}}w_{a_{1}}\left[  ig_{1}^{2}(3\phi_{N}-\sqrt{2}\phi_{C}%
)+2g_{1}\left(  \frac{w_{a_{1}}-iw_{D^{\star0}}}{w_{D^{\star0}}w_{a_{1}}%
}\right)  +ih_{2}(\phi_{N}-\sqrt{2}\phi_{C})-2ih_{3}\phi_{N}\right]  \,.
\end{align}
The decay width for $D_{0}^{\ast0}\rightarrow D^{+}\pi^{-}$ has the same
expression but is multiplied by an isospin factor $2$. The positively charged
scalar state $D_{0}^{\ast+}$ decays into $D^{+}\pi^{0}$ and $D^{0}\pi^{+}$.
The decay width $D_{0}^{\ast+}\rightarrow D^{+}\pi^{0}$ has the same form as
Eq.\ (\ref{d0star}) upon identifying $S,\,P_{1},\,$and $P_{2}$ with
$D_{0}^{\ast+},\,D^{+},$ and $\pi^{0}$. A similar expression holds for the
decay width for $D_{0}^{\ast+}\rightarrow D^{0}\pi^{+}$, where an overall
isospin factor $2$ is also present.

\bigskip

Finally we turn to the scalar state $D_{S0}^{\ast+}$ which decays into
$D^{+}K^{0}\,$and $D^{0}K^{+}$. The decay width $D_{S0}^{\ast+}\rightarrow
D^{+}K^{0}$ is obtained from Eq.\ (\ref{d0star}) upon identifying $S,\,P_{1},$
and $P_{2}$ with $D_{S0}^{\ast+},\,D^{+},$ and $K^{0}$ and upon replacing
$A_{D_{0}^{\ast0}D\pi}\rightarrow A_{D_{S0}^{\ast}DK},\,B_{D_{0}^{\ast0}D\pi
}\rightarrow B_{D_{S0}^{\ast}DK},\,C_{D_{0}^{\ast0}D\pi}\rightarrow
C_{D_{S0}^{\ast}DK},\,E_{D_{0}^{\ast0}D\pi}\rightarrow E_{D_{S0}^{\ast}DK }$,
where:
\begin{align}
A_{D_{S0}^{\ast}DK}  &  =\frac{Z_{K}Z_{D}Z_{D_{S0}^{\ast}}}{\sqrt{2}}%
\lambda_{2}\left[  \phi_{N}+\sqrt{2}(\phi_{S}-\phi_{C})\right]  ,\\
B_{D_{S0}^{\ast}DK}  &  =\frac{Z_{K}Z_{D}Z_{D_{S0}^{\ast}}}{2}\,w_{K_{1}%
}w_{D_{1}}\left[  -\sqrt{2}g_{1}\frac{w_{K_{1}}+w_{D_{1}}}{w_{K_{1}}w_{D_{1}}%
}+\sqrt{2}(g_{1}^{2}-h_{3})\phi_{N}+(g_{1}^{2}+h_{2})(\phi_{S}+\phi
_{C})\right]  \text{ ,}\\
C_{D_{S0}^{\ast}DK}  &  =\frac{Z_{K}Z_{D}Z_{D_{S0}^{\ast}}}{2}\,w_{D_{1}%
}w_{D_{S}^{\ast}}\left[  \sqrt{2}g_{1}\frac{w_{D_{1}}+iw_{D_{S}^{\star}}%
}{w_{D_{1}}w_{D_{S}^{\star}}}-\frac{i}{\sqrt{2}}(g_{1}^{2}+h_{2})\phi
_{N}+i(g_{1}^{2}+h_{2})\phi_{S}+2i(h_{3}-g_{1}^{2})\phi_{C}\right]  \text{
,}\\
E_{D_{S0}^{\ast}DK}  &  =\frac{Z_{K}Z_{D}Z_{D_{S0}^{\ast}}}{2}\,w_{K_{1}%
}w_{D_{S}^{\ast}}\left[  \sqrt{2}g_{1}\frac{w_{K_{1}}-iw_{D_{S}^{\star}}%
}{w_{K_{1}}w_{D_{S}^{\star}}}+\frac{i}{\sqrt{2}}(g_{1}^{2}+h_{2})\phi
_{N}+2i(g_{1}^{2}-h_{3})\phi_{S}-i(g_{1}^{2}+h_{2})\phi_{C}\right]  \text{ .}%
\end{align}
The decay of the scalar state $D_{S0}^{\ast}$ into $D^{0}K^{+}$ has an
analogous analytic expression.

\subsection{Decay widths of charmed vector mesons}

\label{app2}The neutral state $D^{\ast0}$ decays into $D^{0}\pi^{0}$. The
corresponding expression is:%

\begin{equation}
\Gamma_{V\rightarrow P_{1}P_{2}}=\frac{1}{24\pi}\left[  \frac{(m_{V}%
^{2}-m_{P_{1}}^{2}-m_{P_{2}}^{2})^{2}-4m_{P_{1}}^{2}m_{P_{2}}^{2}}{4m_{V}^{4}%
}\right]  ^{3/2}(A_{D^{\ast}D\pi}-B_{D^{\ast}D\pi}+C_{D^{\star}D\pi}%
\,m_{V}^{2})^{2}\;, \label{DsDpiQ}%
\end{equation}
where
\begin{align}
A_{D^{\ast}D\pi}  &  =\frac{i}{2}Z_{\pi}Z_{D}\left[  g_{1}+\sqrt{2}w_{D_{1}%
}(h_{3}-g_{1}^{2})\phi_{C}\right]  \;,\\
B_{D^{\ast}D\pi}  &  =-\frac{i}{4}Z_{\pi}Z_{D}\left[  2g_{1}-w_{a_{1}}%
(3g_{1}^{2}+h_{2}-2h_{3})\phi_{N}+\sqrt{2}w_{a_{1}}(g_{1}^{2}+h_{2})\phi
_{C}\right]  \;,\\
C_{D^{\ast}D\pi}  &  =\frac{i}{2}Z_{\pi}Z_{D}w_{a_{1}}w_{D_{1}}g_{2}\;.
\end{align}
and where $V,$ $P_{1},$and $P_{2}$ refer to $D^{\ast0},\,D^{0},\,$and $\pi
^{0}$.

For the decay $D^{\ast+}\rightarrow D^{+}\pi^{0}$ Eq.\ (\ref{DsDpiQ}) holds
upon multiplication by an isospin factor $2.$ .

\subsection{Decay widths of charmed axial-vector mesons}

\label{app3} The neutral nonstrange axial-vector meson $D_{1}^{0}$ decays into
$D^{\ast0}\pi^{0},\,D^{\ast+}\pi^{-},\,D^{0}\pi^{0}\pi^{0},\,D^{0}\pi^{+}%
\pi^{-}$, and $D^{-}\pi^{+}\pi^{0}$. The decay width for $D_{1}^{0}\rightarrow
D^{\ast0}\pi^{0}$ is
\begin{equation}
\Gamma_{A\rightarrow VP}=\frac{K(m_{A},m_{V},m_{P})}{12\pi m_{A}^{2}}\left[
\left\vert h_{D_{1}D^{\ast}\pi}^{\mu\nu}\right\vert ^{2}-\frac{\left\vert
h_{D_{1}D^{\ast}\pi}^{\mu\nu}k_{A,\mu}\right\vert ^{2}}{m_{A}^{2}}%
-\frac{\left\vert h_{D_{1}D^{\ast}\pi}^{\mu\nu}k_{V,\nu}\right\vert ^{2}%
}{m_{V}^{2}}+\frac{\left\vert h_{D_{1}D^{\ast}\pi}^{\mu\nu}k_{A,\mu}%
\,k_{V,\nu}\right\vert ^{2}}{m_{A}^{2}m_{V}^{2}}\right]  \text{ ,}
\label{d1dspionQ}%
\end{equation}
where $A,\,V,\,$and $P$ refer to the mesons $D_{1}^{0}$, $D^{\ast0},\,\pi^{0}%
$and $h_{D_{1}D^{\ast}\pi}^{\mu\nu}$ is
\begin{equation}
h_{D_{1}D^{\ast}\pi}^{\mu\nu}=i\left\{  A_{D_{1}D^{\ast}\pi}g^{\mu\nu
}+B_{D_{1}D^{\ast}\pi}[k_{V}^{\mu}k_{P}^{\nu}+k_{P}^{\mu}k_{A}^{\nu}%
-(k_{A}\cdot k_{P})g^{\mu\nu}-(k_{V}\cdot k_{P})g^{\mu\nu}]\right\}  \text{ ,}
\label{hD1Dspi}%
\end{equation}
with
\begin{align}
A_{D_{1}D^{\ast}\pi}  &  =\frac{i}{\sqrt{2}}Z_{\pi}(g_{1}^{2}-h_{3})\phi
_{C}\;\text{,}\label{AD1Dpi2}\\
B_{D_{1}D^{\ast}\pi}  &  =\frac{i}{2}Z_{\pi}g_{2}w_{a_{1}}\text{ .}
\label{BD1Dpi2}%
\end{align}
The quantities $k_{A}^{\mu}=(m_{A},\mathbf{0})$, $k_{V}^{\mu}=(E_{V}%
,\mathbf{k})$, and $k_{P}^{\mu}=(E_{P},-\mathbf{k})$ are the four-momenta of
$D_{1}^{0}$, $D^{\ast0}$, and $\pi^{0}$ in the rest frame of $D_{1}^{0}$,
respectively. The following kinematic relations hold:
\begin{align*}
&  k_{V}\cdot k_{P}=\frac{m_{A}^{2}-m_{V}^{2}-m_{P}^{2}}{2}\,,{}\\
&  k_{A}\cdot k_{V}=m_{A}E_{V}=\frac{m_{A}^{2}+m_{V}-m_{P}^{2}}{2}\,,\\
&  k_{A}\cdot k_{P}=m_{A}E_{P}=\frac{m_{A}^{2}-m_{V}+m_{P}^{2}}{2}\,.
\end{align*}
The terms entering in Eq.\ (\ref{d1dspionQ}) are given by%

\begin{align}
\left\vert h_{D_{1}D^{\ast}\pi}^{\mu\nu}\right\vert ^{2}  &  =4A_{D_{1}%
D^{\ast}\pi}^{2}+B_{D_{1}D^{\ast}\pi}^{2}\left[  m_{V}^{2}m_{P}^{2}+m_{A}%
^{2}m_{P}^{2}+2(k_{V}\cdot k_{P})^{2}+2(k_{A}\cdot k_{P})^{2}+6(k_{V}\cdot
k_{P})(k_{A}\cdot k_{P})\right] \nonumber\\
&  -6A_{D_{1}D^{\ast}\pi}B_{D_{1}D^{\ast}\pi}(k_{V}\cdot k_{P}+k_{A}\cdot
k_{P})\text{,} \label{hD1dspQ1}%
\end{align}%
\begin{align}
\left\vert h_{D_{1}D^{\ast}\pi}^{\mu\nu}k_{A,\mu}\right\vert ^{2}  &
=A_{D_{1}D^{\ast}\pi}^{2}m_{A}^{2}+B_{D_{1}D^{\ast}\pi}^{2}\left[  (k_{A}\cdot
k_{V})^{2}m_{P}^{2}+(k_{V}\cdot k_{P})^{2}m_{A}^{2}-2(k_{A}\cdot k_{V}%
)(k_{A}\cdot k_{P})(k_{V}\cdot k_{P})\right] \nonumber\\
&  +2A_{D_{1}D^{\ast}\pi}B_{D_{1}D^{\ast}\pi}\left[  (k_{A}\cdot k_{V}%
)(k_{A}\cdot k_{P})-(k_{V}\cdot k_{P})m_{A}^{2}\right]  \text{,}%
\end{align}%
\begin{align}
\left\vert h_{D_{1}D^{\ast}\pi}^{\mu\nu}k_{V,\nu}\right\vert ^{2}  &
=A_{D_{1}D^{\ast}\pi}^{2}m_{V}^{2}+B_{D_{1}D^{\ast}\pi}^{2}\left[  (k_{A}\cdot
k_{V})^{2}m_{P}^{2}+(k_{A}\cdot k_{P})^{2}m_{V}^{2}-2(k_{A}\cdot k_{V}%
)(k_{A}\cdot k_{P})(k_{V}\cdot k_{P})\right] \nonumber\\
&  +2A_{D_{1}D^{\ast}\pi}B_{D_{1}D^{\ast}\pi}\left[  (k_{A}\cdot k_{V}%
)(k_{V}\cdot k_{P})-(k_{A}\cdot k_{P})m_{V}^{2}\right]  \text{ ,}%
\end{align}%
\[
\left\vert h_{D_{1}D^{\ast}\pi}^{\mu\nu}k_{A,\mu}k_{V,\nu}\right\vert
^{2}=[A_{D_{1}D^{\ast}\pi}(k_{A}\cdot k_{V})]^{2}%
\]
\newline with $E_{V}=\sqrt{K^{2}(m_{A},m_{V},m_{P})+m_{V}^{2}}$ and
$E_{P}=\sqrt{K^{2}(m_{A},m_{V},m_{P})+m_{P}^{2}}$.

The decay width for $D_{1}^{0}\rightarrow D^{\ast+}\pi^{-}$ is still given by
Eq.\ (\ref{d1dspionQ}) upon substituting the fields and upon multiplication by
an isospin factor $2.$

Now we are turning to the three body-decay of the axial-vector meson
$D_{1}^{0}$, which decays into three pseudoscalar mesons ($D\pi\pi$). First,
the decay width for $D_{1}^{0}\rightarrow D^{0}\pi^{0}\pi^{0}$ can be obtained
as
\begin{equation}
\Gamma_{A\rightarrow P_{1}P_{2}P_{3}}=\frac{2}{96\,(2\pi)^{3}\,M^{3}}%
\int_{(m_{1}+m_{2})^{2}}^{(M-m_{3})^{2}}\int_{(m_{23})_{\min}}^{(m_{23}%
)_{\max}}\left[  -\left\vert h_{D_{1}D\pi\pi}^{\mu}\right\vert ^{2}%
+\frac{\left\vert h_{D_{1}D\pi\pi}^{\mu}k_{\mu}\right\vert ^{2}}{M^{2}%
}\right]  \,dm_{23}^{2}\,dm_{12}^{2}\>, \label{d1d2pionQ}%
\end{equation}
where $h_{D_{1}D\pi\pi}^{\mu}$ is the vertex following from the relevant part
of the Lagrangian
\begin{equation}
h_{D_{1}D\pi\pi}^{\mu}=-\left[  A_{D_{1}D\pi\pi}g^{\mu}k_{1}^{\mu}%
+B_{D_{1}D\pi\pi}k_{2}^{\mu}\right]  \text{ ,} \label{hD1Dpipi}%
\end{equation}
with the coefficients
\begin{align}
A_{D_{1}D\pi\pi}  &  =\frac{1}{4}Z_{\pi}^{2}\,Z_{D}\,w_{D_{1}}(g_{1}%
^{2}+2h_{1}+h_{2})\;,\label{AD1Dpi}\\
B_{D_{1}D\pi\pi}  &  =\frac{1}{4}Z_{\pi}^{2}\,Z_{D}\,w_{a_{1}}(3g_{1}%
^{2}+h_{2}-2h_{3})\text{ ,} \label{BD1Dpi}%
\end{align}
and $k^{\mu}=(M,\mathbf{0})$, $k_{1}^{\mu}=(E_{P_{1}},\mathbf{k})$,
$k_{2}^{\mu}=(E_{P_{2}},-\mathbf{k})$, and $k_{3}^{\mu}=(E_{P_{3}}%
,-\mathbf{k})$ are the four-momenta of $D_{1}^{0}$, $D^{0}$, $\pi^{0}$, and
$\pi^{0}$ in the rest frame of $D_{1}^{0}$, respectively. Using the following
kinematic relations
\begin{align*}
&  k_{1}\cdot k_{2}=\frac{m_{12}^{2}-m_{1}^{2}-m_{2}^{2}}{2}\,,\\
&  k\cdot k_{1}=m_{1}^{2}+\frac{m_{12}^{2}-m_{1}^{2}-m_{2}^{2}}{2}%
+\frac{m_{13}^{2}-m_{1}^{2}-m_{3}^{2}}{2}\,,\\
&  k\cdot k_{2}=m_{2}^{2}+\frac{m_{12}^{2}-m_{1}^{2}-m_{2}^{2}}{2}%
+\frac{m_{23}^{2}-m_{2}^{2}-m_{3}^{2}}{2}\,,
\end{align*}
where the quantities $M,\,m_{1},\,m_{2},\,m_{3}$ are the masses of $D_{1}%
^{0},\,D^{0},\,\pi^{0},\,\pi^{0}$, respectively. The decay of the scalar state
$D_{1}^{+}\rightarrow D^{+}\pi^{0}\pi^{0}$ has an analogous analytic
expression. The decay of the scalar states $D_{1}^{0}\rightarrow D^{0}\pi
^{+}\pi^{-}$ and $D_{1}^{+}\rightarrow D^{0}\pi^{+}\pi^{-}$ have the same
formula as presented in Eq.\ (\ref{d1d2pionQ}) but multiplied by an isospin
factor $2$ as given by the Lagrangian (\ref{lag}). \newline

The decay width $D_{1}^{0}\rightarrow D^{+}\pi^{-}\pi^{0}$ reads
\begin{equation}
\Gamma_{A\rightarrow P_{1}P_{2}P_{3}}=\frac{F_{D_{1}D\pi\pi}^{2}}%
{96\,(2\pi)^{3}\,M^{3}}\int_{(m_{1}+m_{2})^{2}}^{(M-m_{3})^{2}}\int
_{(m_{23})_{\min}}^{(m_{23})_{\max}}\left[  \frac{\left\vert k\cdot
k_{2}-k\cdot k_{3}\right\vert ^{2}}{M^{2}}-(m_{2}^{2}+m_{3}^{2}+2k_{2}\cdot
k_{3})\right]  \,dm_{23}^{2}\,dm_{12}^{2}\>,
\end{equation}
where the quantities $A,\,P_{1},\,P_{2},\,$and $P_{3}$ refer to the fields
$D_{1}^{0},\,D^{+},\,\pi^{-},$ and $\pi^{0}$, respectively. Moreover:
\begin{equation}
F_{D_{1}D\pi\pi}=\frac{\sqrt{2}}{4}Z_{\pi}^{2}\,Z_{D}\,w_{a_{1}}(g_{1}%
^{2}-h_{2}-2h_{3})\text{ ,} \label{FD1Dpi}%
\end{equation}
and
\begin{align*}
&  k_{2}\cdot k_{3}=\frac{m_{23}^{2}-m_{2}^{2}-m_{3}^{2}}{2}\,,\\
&  k\cdot k_{3}=m_{3}^{2}+\frac{m_{13}^{2}-m_{1}^{2}-m_{3}^{2}}{2}%
+\frac{m_{23}^{2}-m_{2}^{2}-m_{3}^{2}}{2}\,.
\end{align*}
The charged state $D_{1}^{+}$ decays into $D^{\ast+}\pi^{0},\,D^{\ast0}\pi
^{+},\,D^{+}\pi^{0}\pi^{0},\,D^{+}\pi^{+}\pi^{-},$ and $D^{0}\pi^{0}\pi^{+}$.
Analogous expressions hold.

As a last step we turn to the strange-charmed axial-vector state $D_{S1}^{+}$.
It decays predominantly into the channels $D^{\ast+}K^{0}$ and $D^{\ast0}%
K^{+}$. The formula for the decay width $\Gamma_{D_{S1}^{+}\rightarrow
D^{\ast+}K^{0}}$ and $\Gamma_{D_{S1}^{+}\rightarrow D^{\ast0}K^{0}}$ is as in
Eq.\ (\ref{d1dspionQ}) upon replacing the fields:
\begin{equation}
h_{D_{S1}D^{\ast}K}^{\mu\nu}=i\left\{  A_{D_{S1}D^{\ast}K}g^{\mu\nu}%
+B_{D_{S1}D^{\ast}K}[P_{1}^{\mu}P_{2}^{\nu}+P_{2}^{\mu}P^{\nu}-(P\cdot
P_{2})g^{\mu\nu}-(P_{1}\cdot P_{2})g^{\mu\nu}]\right\}  \text{ ,}
\label{hDs1Dsk}%
\end{equation}
where%
\begin{align}
&  A_{D_{S1}D^{\ast}K}=\frac{i}{4}Z_{K}\left[  g_{1}^{2}(\sqrt{2}\phi
_{N}-2\phi_{S}-4\phi_{C})+h_{2}(\sqrt{2}\phi_{N}-2\phi_{S})+4h_{3}\phi
_{C}\right]  ,\\
&  B_{D_{S1}D^{\ast}K}=-\frac{i}{\sqrt{2}}Z_{K}\,g_{2}w_{K_{1}}.
\end{align}

\end{document}